\documentclass[twocolumn,english,superscriptaddress,amsmath,floatfix,longbibliography,floats,prx]{revtex4-1}

\usepackage[colorlinks=true,urlcolor=blue,citecolor=blue,linkcolor=blue]{hyperref} 
\usepackage[T1]{fontenc}
\usepackage[utf8]{inputenc}
\usepackage{amssymb}
\usepackage{graphicx}
\usepackage{amsmath,color}
\usepackage{mathrsfs}
\usepackage{float}
\usepackage{indentfirst}
\usepackage{txfonts}
\usepackage[normalem]{ulem}
\usepackage[table]{xcolor}
\usepackage{array,ragged2e}
\usepackage{grffile}
\usepackage{multirow}
\usepackage{algpseudocode}
\usepackage{epstopdf}
\makeatletter

\newcommand{\RNum}[1]{\uppercase\expandafter{\romannumeral #1\relax}}
\newcommand{\rNum}[1]{\romannumeral #1}

\makeatother

\usepackage{babel}

\begin{document}

\hyphenpenalty=5000

\tolerance=1000

\title{Tensor network study of the spin-1/2 square-lattice $J_1$-$J_2$-$J_3$ model: incommensurate spiral order, mixed valence-bond solids, and multicritical points}
\author{Wen-Yuan Liu}
\affiliation{Division of Chemistry and Chemical Engineering, California Institute of Technology, Pasadena, California 91125, USA}
\author{Didier Poilblanc}
\affiliation{Laboratoire de Physique Th\'eorique, C.N.R.S. and Universit\'e de Toulouse, 31062 Toulouse, France}
  \author{Shou-Shu Gong}
 \affiliation{School of Physical Sciences, Great Bay University, Dongguan 523000, China, and \\
 Great Bay Institute for Advanced Study, Dongguan 523000, China}
 \author{Wei-Qiang Chen}
 \affiliation{Shenzhen Institute for Quantum Science and Engineering and Department of Physics, Southern University of Science and Technology, Shenzhen 518055, China}
 \affiliation{Shenzhen Key Laboratory of Advanced Quantum Functional Materials and Devices, Southern University of Science and Technology, Shenzhen 518055, China}
 \author{Zheng-Cheng Gu}
 \affiliation{Department of Physics, The Chinese University of Hong Kong, Shatin, New Territories, Hong Kong, China}

\date{\today }

\begin{abstract}

We use the finite projected entangled pair state (PEPS) method to investigate the global phase diagram of the spin-1/2 square-lattice $J_1$-$J_2$-$J_3$ antiferromagnetic (AFM) Heisenberg model. The ground-state phase diagram is established with a rich variety of phases: N\'eel AFM, gapless quantum spin liquid, valence-bond solid (VBS), stripe AFM, and incommensurate spiral phases. The nature of the VBS region is revealed, which contains a plaquette VBS and a mixed columnar-plaquette VBS, with the emergence of short-range incommensurate spin correlation in some region. The long-range incommensurate magnetic phase is also explicitly characterized as a planar spiral with incommensurate spatial periodicities. Most interestingly, there exists several multicritical points connecting different phases. These findings elucidate the true nature of the long-standing square-lattice $J_1$-$J_2$-$J_3$ antiferromagnet at zero-temperature. Our results also pave the way to accurately simulate complex two-dimensional quantum spin systems that may host nonuniform features by means of the finite PEPS.   

\end{abstract}

\date{\today}
\maketitle

\section{Introduction}

\begin{figure}[htbp]
 \centering
 \includegraphics[width=3.4in]{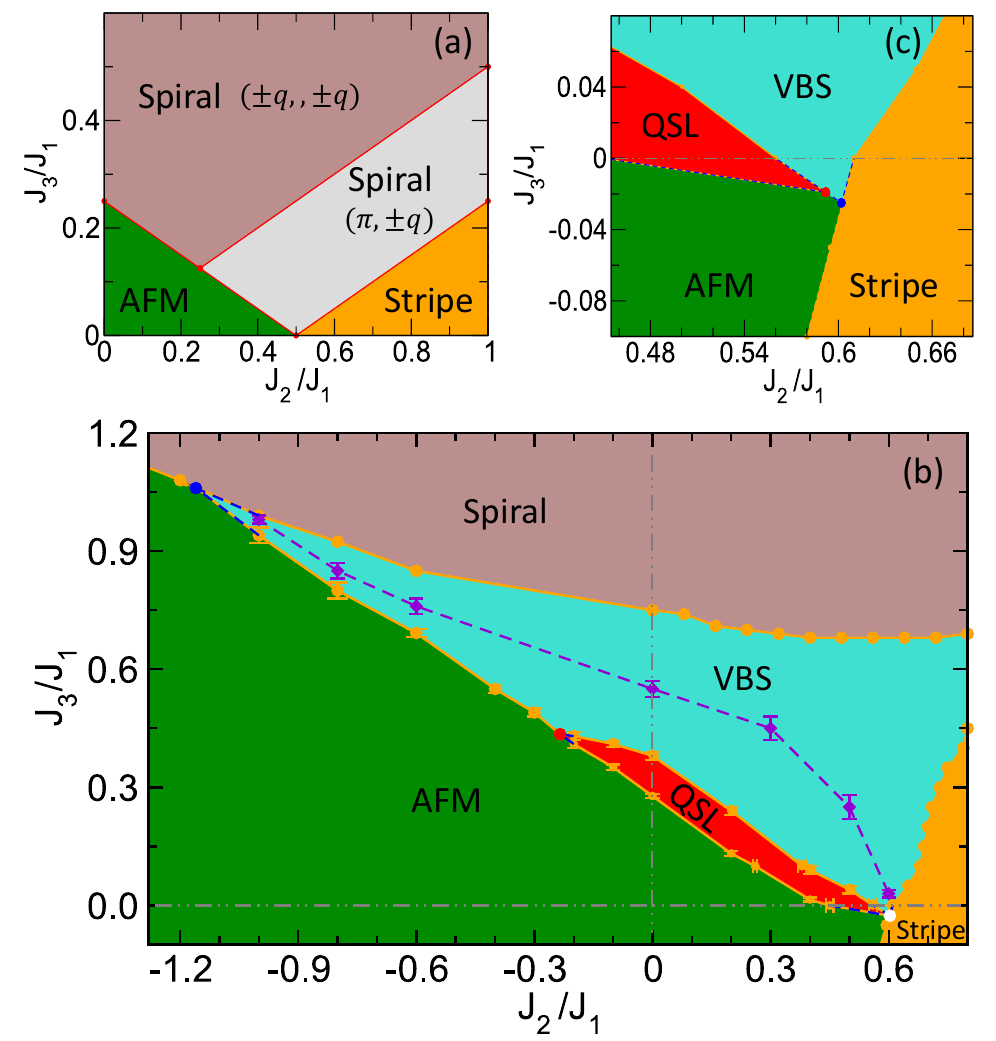}
 \caption{
 (a) Classical phase diagram for the ground state of the $J_1$-$J_2$-$J_3$ model: the AFM phase ordered at $(\pi,\pi)$, the stripe phase ordered at $(0,\pi)$ or $(\pi,0)$, a spiral phase ordered at $(\pm q,\pm q)$, and another spiral phase ordered at $(\pi,\pm q)$ or $(\pm q, \pi)$. (b) Quantum phase diagram for the ground state of the spin-1/2 square-lattice $J_1$-$J_2$-$J_3$ AFM model:  the AFM phase ordered at $(\pi,\pi)$, the stripe phase ordered at $(0,\pi)$ or $(\pi,0)$, the spiral phase ordered at $(\pm q,\pm q)$, the gapless QSL phase (red) and the VBS phase. Tricritical points (end points) are shown by red (blue) dots. The AFM-spiral, AFM-stripe, VBS-spiral and VBS-stripe transitions are first-order, and other phase transitions are continuous. The phase transition lines are determined via finite size extrapolations. In the VBS region, below the violet dashed line is a plaquette VBS  and above is a mixed columnar-plaquette VBS. The region in the right corner might contain a quadruple point (the white dot) connecting the  VBS, QSL, AFM and stripe phases, or  two multicritical points shown in (c) (assuming the gapless QSL does not touch the stripe phase). Grey dashed-dotted lines marking  $J_2=0$ and $J_3=0$ are shown in (b) and (c).  
}
 \label{fig:J1J2J3phaseDiagram}
 \end{figure}

The spin-1/2 $J_1$-$J_2$-$J_3$ antiferromagnetic (AFM)  Heisenberg model on the square lattice is one of the paradigmatic prototypes to study frustrated quantum magnets. This model has attracted a lot of interests after the application of  Anderson’s resonating valence-bond (RVB) theory to high-temperature superconductivity~\cite{RVB}. 
The Hamiltonian of this system is described as follow:
\begin{equation}
H=J_1\sum_{\langle i,j \rangle}\mathbf{S_i}\cdot\mathbf{S_j}+J_2\sum_{\langle\langle
i,j\rangle\rangle}\mathbf{S_i}\cdot\mathbf{S_j}+J_3\sum_{\langle\langle\langle
i,j\rangle\rangle\rangle}\mathbf{S_i}\cdot\mathbf{S_j},\quad 
\label{model}
\end{equation}
where $J_1$, $J_2$, $J_3$ denote the first-, second- and third-nearest neighbour couplings respectively, and the summations run over all corresponding spin pairs. For the positive $J_i$ couplings, the three kinds of interactions compete with each other and thus leads to intractable difficulties for analytic and numerical studies.
The classical phase diagram obtained by spin-wave theory contains four phases:  (\rNum{1}) a N\'eel AFM phase ordered at wave vector ${\bf k_0}=(\pi,\pi)$, (\rNum{2}) a stripe AFM phase ordered at wave vector ${\bf k_x}=(\pi,0)$ or ${\bf k_y}=(0,\pi)$, (\rNum{3}) a spiral phase at ${\bf Q_x}=(\pm q,\pi)$ or ${\bf Q_y}=(\pi, \pm q)$  with $\cos{q}=(2J_2-J_1)/4J_3$, and (\rNum{4}) another spiral phase at  ${\bf Q}=(\pm q,\pm q)$ with $\cos{q}=-J_1/(2J_2+4J_3)$~\cite{seriesexpansion1989,spinwave1990}, as shown in Fig.~\ref{fig:J1J2J3phaseDiagram}(a). The N\'eel AFM and the spiral  $(\pm q,\pm q)$ phases are separated by a classical critical line $(J_2+2J_3)=J_1/2$. 
However, the quantum phase diagram with spin-$1/2$ is not fully understood.

For spin-$1/2$, it has been established that at small $J_2$ and $J_3$ couplings, the model possesses a N\'eel AFM order~\cite{chandra1988,ed2,spinwave1990,spinwave1991}. 
At larger $J_2$ and $J_3$, the corresponding stripe AFM and spiral orders will develop~\cite{LargeN1991}. 
The controversy exists in the intermediate region of $J_2$ and $J_3$, especially along the classical critical line. 
Some theories strongly suggest that the  combined effect of enhanced quantum fluctuations and frustration could destroy the long-range orders and stabilize paramagnetic states along the critical line, including the spin-wave theories~\cite{chandra1988,spinwave1990,spinwave1991}, renormalization group (RG) analysis of the non-linear $\sigma$ model~\cite{RG1988}, series expansions~\cite{seriesexpansion1989}, and momentum-shell RG calculation~\cite{momentumshell1991}, etc. 
There are also other theories supporting the existence of paramagnetic states but for the $J_2$ and $J_3$ couplings shifted to larger values with respect to the classical critical line by quantum fluctuations~\cite{LargeN1991,ferrer1993}.

From the combined analyses of different studies, it seems most likely there exists an intermediate nonmagnetic region. 
Nevertheless, the nature of the nonmagnetic region is far from clear. 
While the large-$N$ and series expansion results predict that this intermediate state is spontaneously dimerized~\cite{LargeN1991,seriesexpansion1989}, the spin-wave theory suggests a spin liquid state~\cite{chandra1988,spinwave1990,spinwave1991}.  
For the $J_1$-$J_3$ model with $J_2=0$, a Monte Carlo study of the classical limit supplemented by analytical arguments on the role of quantum fluctuations supports the emergence of a valence-bond solid (VBS) or a $Z_2$ spin liquid between the N\'eel AFM and spiral phases~\cite{classicalJ1J3}. 
While exact diagonalization calculations on small clusters suggest a VBS state with incommensurate short-range spin correlation~\cite{ed1996}, the early density matrix renormalization group (DMRG) results may support a gapped spin liquid~\cite{DMRGJ1J3}.  
For $J_2\neq 0$, a short-range valence bond study finds a plaquette VBS state along the line with $J_2+J_3=J_1/2$ (where the description in terms of nearest-neighbor singlet coverings is excellent)~\cite{PVB4}. 
Later, a mixed columnar-plaquette VBS state was also proposed~\cite{mixVBS2009}, which possesses a long-range plaquette order but breaks the isotropy between the $x$ and $y$ directions, supporting the findings in a  previous study~\cite{GFMC2000}.
 Furthermore, exact diagonalizations also show that quantum fluctuations could lead to new quantum phases~\cite{ed2010}.

The special case of the $J_1$-$J_2$-$J_3$ model with $J_3=0$, i.e., the $J_1$-$J_2$ model, is of great interest in the field of quantum magnetism, which has been intensively studied for more than three decades. The $J_1$-$J_2$ model exhibits a N\'eel AFM phase in the region $0\leq J_2/J_1\lesssim 0.45$ and a stripe AFM phase for $ J_2/J_1\geq 0.61$. The intermediate nonmagnetic region has been investigated by many different methods~\cite{chandra1988,ed1,ed2,CVB1,ed7,CVB2,ed3,ivanov1992,ed5,ed4,PVB1,CVB3,PVB2,VMC2001,gQSL2,PVB3,sirker2006,schmalfu2006,PVB4,PVB5,PVB6,PVB7,murg2009,beach2009,ed6,PVB8,jiang2012,mezzacapo2012,WangRVB,hu2013,PVB9,YangJ1J2,gong2014,chou2014,morita2015,morita2015,richter2015,wang2016,poilblanc2017,wang2018,CVB4,liu2018,poilblanc2019,hasik2021,ferrari2020,liuQSL,nomura2020}, and the existence of two successive gapless quantum spin liquid (QSL) and VBS phases lying between the N\'eel and stripe phase is being supported by more recent works~\cite{wang2018,ferrari2020,liuQSL,nomura2020}. 
In particular, the finite PEPS method provides very solid large-size results and makes it possible to extract critical exponents to understand the physical nature of the AFM-QSL and QSL-VBS phase transitions~\cite{liuQSL}, which indicates an intrinsic relation between the gapless QSL and deconfined quantum critical point (DQCP)~\cite{DQCP1}.

Very recently, a tensor network study has also revealed the exotic properties of the $J_1$-$J_2$-$J_3$ model~\cite{liu2022emergence}. 
Specifically, the highly accurate tensor network results from finite and infinite PEPS simulations not only identify the extended QSL and VBS phases at finite $J_3$, but also demonstrate the intrinsic relation between the gapless QSL and DQCP, offering a fantastic scenario to understand the two exotic quantum phenomena. The existence of QSL and VBS phases in the pure $J_1$-$J_3$ model, i.e., $J_2=0$, is also supported by recent DMRG calculations~\cite{wu2022phase}.

However, because of the highly complex competition of $J_1$, $J_2$ and $J_3$ interactions, the global phase diagram of the $J_1$-$J_2$-$J_3$ model remains enigmatic, particularly in the regions that may involve nonuniform physical properties like the incommensurate spin correlations. Thanks to the advancement of tensor network method, we can now study this model in a precise way. 
One approach to study this model is based on the widely used infinite PEPS (iPEPS) ansatz~\cite{liu2022emergence}. Although it has a presumed unit cell, the nonuniform properties can still be investigated by systematically enlarging the  size of unit cell. There have been remarkable progresses in correlated electron models along this direction~\cite{corboz2011stripes,corboz2014competing,zheng2017stripe}. In this paper, we shed new light on the ground-state phase diagram of the $J_1$-$J_2$-$J_3$ model by using an alternative approach, namely, the finite PEPS ansatz, which is well-suited for describing nonuniform features with site-independent tensors. The finite PEPS works very well in the scheme combined with variational Monte Carlo, and the full details of the algorithm can be  found in the preceding publications~\cite{liu2017,liufinitePEPS,liuQSL,liu2022emergence}. The obtained phase diagram is presented in Fig.~\ref{fig:J1J2J3phaseDiagram}(b).

Essentially, this paper is  an extension of our  previous work on the $J_1$-$J_2$-$J_3$ model~\cite{liu2022emergence}, but emphasizing different aspects.  In the previous work, we discovered a large region of gapless QSL between the N\'eel AFM and VBS phases. By tuning coupling constants we found that the extension of the gapless QSL decreases and eventually merges into a line of direct continuous transition between the N\'eel AFM and VBS phases, showing a direct connection between gapless QSL and DQCP. With further analyses of the AFM-VBS, AFM-QSL and QSL-VBS transitions, we explicitly demonstrated a new scenario to understand gapless QSL and DQCP, suggesting they are described by a unified quantum field theory.  The gapless QSL and DQCP physics actually is only a part of the $J_1$-$J_2$-$J_3$ model, and here we turn to reveal its global ground state phase diagram. Specifically, we comprehensively investigate its nonuniform features including incommensurate short-range and long-range spiral properties.
We also find a novel mixed valence-bond solid phase as well as several multicritical points.

Rest of the paper is organized as follows. 
In Sec.~\RNum{2}, we first present the results obtained at $J_2=0$. By computing spin and dimer order parameters, as well as ground state energies, we provide a complete phase diagram of the $J_1$-$J_3$ model, and focus on the emergence of incommensurate spiral spin correlations upon increasing $J_3$. Then, we consider an intermediate $J_2\neq 0$ term to further analyze the properties of the VBS phase. Next we consider a large negative $J_2$, which can suppress the intermediate VBS phase and eventually lead to a direct first-order AFM-spiral phase transition. Finally, we consider the $J_3<0$ case to obtain the complete phase diagram, where the gapless QSL and VBS phases eventually disappear, and instead a direct AFM-stripe transition occurs. In Sec.~\RNum{3}, we discuss the existing results and obtain a refined description of the phase diagram of the $J_1$-$J_2$-$J_3$ model.   
Finally, three appendices are provided, including the convergence with bond dimension, the ED results of a $4\times 4$ cluster, and a refined analysis of the spin structure factor in the mixed VBS phase.

\section{results}

We use the recently developed finite PEPS method to perform all the calculations~\cite{liu2017,liufinitePEPS}. The finite PEPS method works very well in the scheme of variational Monte Carlo approach, where the summation of physical degrees of freedom is replaced by Monte Carlo sampling.
Thus, one only needs to deal with the single-layer tensor networks at a computational cost scaling ${\cal{O}}(D^6)$, where $D$ is the tensor bond dimension. 
Such an approach has been successfully applied to solve very challenging quantum many-body problems~\cite{liuQSL,liu2022emergence,j1xj1yj2,liu2023deconfined}.  
In Ref.~\cite{liu2022emergence}, we have demonstrated that the simulations with $D=8$ can obtain well converged results for the $J_1$-$J_2$-$J_3$ model up to the $20\times 28$ system size. 
Here we have checked the $D$-convergence on the open $16\times 16$ cluster at $(J_2,J_3)=(-1, 0.8)$ and find that $D=8$ can also provide converged energy and order parameters, as shown in Appendix~\ref{app:converge}. Therefore, we use $D=8$ for all the calculations and we set $J_1=1$ as the energy unit unless otherwise specified.

\subsection {$J_2 =0$}

\subsubsection{Magnetic order parameters}

\begin{figure}[tbp]
 \centering
 \includegraphics[width=3.4in]{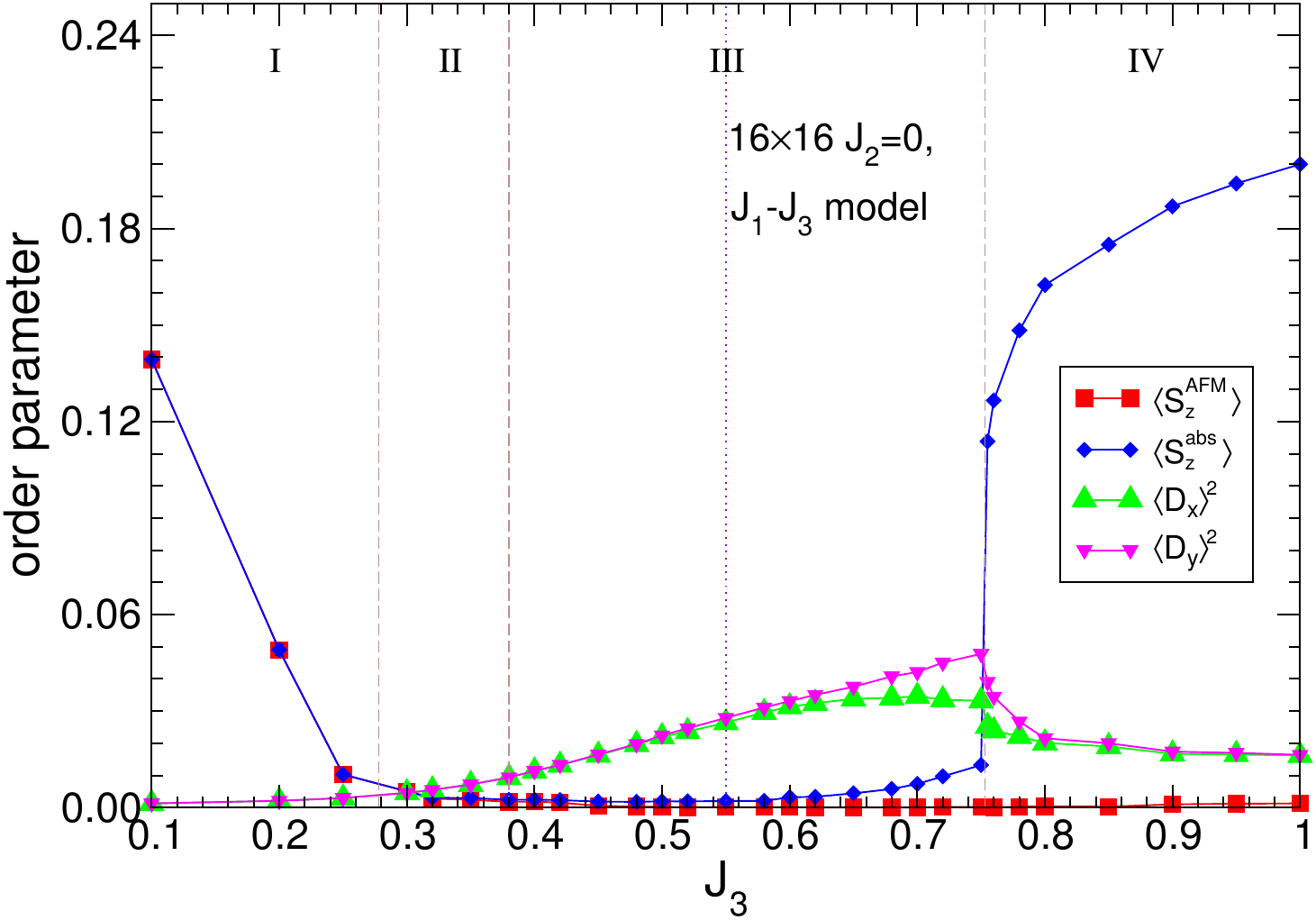}
 \caption{The variation of the local order parameters with respect to $J_3/J_1$ (for $J_2=0$) on a $16\times 16$ system, including  $\langle S_z^{\rm AFM}\rangle $, $\langle S_z^{\rm abs}\rangle$ and $\langle D_x\rangle^2$ and $\langle D_y\rangle^2$.  Phase boundaries between the different AFM (I), QSL (II), VBS (III) and spiral (IV) phases are denoted by vertical dashed lines, located at $J_3=0.28$, $0.38$ and $0.752$, respectively. The violet dotted line in the VBS phase (III) denotes the transition between the plaquette VBS and the mixed-VBS phase. }
 \label{fig:16x16J1J3localOrder}
 \end{figure}

Through detailed computations, we find four phases by varying $J_3$:  (\RNum{1}) the N\'eel AFM  phase ordered at ${\bf k_0}=(\pi,\pi)$, (\RNum{2}) the gapless QSL phase, (\RNum{3}) the VBS phase, and (\RNum{4}) the long-range spiral ordered phase. These four phases have been displayed in Fig.~\ref{fig:16x16J1J3localOrder} along the $J_3$ axis. 
Note that, when $J_3$ is sufficiently large, phase (\RNum{4}) evolves continuously to approach the commensurate spiral  state ordered at ${\bf k_1}=(\pm\pi/2,\pm\pi/2)$.

The gapless spin liquid phase and associated AFM-QSL and QSL-VBS phase transitions have been studied in a previous finite PEPS simulation work~\cite{liu2022emergence}. 
Now we focus on the VBS-spiral transition. 
We define two local order parameters to detect long-range spin ordered phases:
\begin{align}
&\langle S_z^{\rm abs}\rangle =\frac{1}{L^2}\sum_{i_x,i_y}|\langle S^z_{i_x,i_y}\rangle|, \\
& \langle S_z^{\rm AFM}\rangle =\frac{1}{L^2}\sum_{i_x,i_y}(-1)^{i_x+i_y}\langle S^z_{i_x,i_y}\rangle.
\end{align}
Since $\langle S_z^{\rm abs}\rangle$ is the average of the absolute value of the onsite $\langle S^z_{i_x,i_y}\rangle$, it can distinguish magnetic phases from nonmagnetic ones. Note $\langle S_z^{\rm AFM}\rangle$ is the standard N\'eel AFM order parameter, which can distinguish the N\'eel AFM phase from other phases including magnetic ones. These two order parameters should be zero theoretically on finite-size systems because of the SU(2) symmetry, but it is expected that for magnetic phases the corresponding ground state may break SU(2) symmetry and show nonzero values for  $\langle S_z^{\rm abs}\rangle$ or  $\langle S^z_{i_x,i_y}\rangle$ if the system size becomes large enough.

In Fig.~\ref{fig:16x16J1J3localOrder}, we show the variation of  $\langle S_z^{\rm abs}\rangle$ and  $\langle S_z^{\rm AFM}\rangle$ with increasing $J_3$ on the $16\times 16$ system size. Indeed, in the N\'eel AFM phase, $\langle S_z^{\rm abs}\rangle$ and $\langle S_z^{\rm AFM}\rangle$ are equal and finite.  After entering into the nonmagnetic QSL and VBS phases, SU(2) symmetry is almost restored, signaled by the much smaller values of $\langle S_z^{\rm abs}\rangle$ and $\langle S_z^{\rm AFM}\rangle$ close to zero. When approaching the  spiral phase, SU(2) symmetry is broken again and $\langle S_z^{\rm abs}\rangle$ enhances sharply, indicating a first-order phase transition to a magnetic phase at $J_3\simeq 0.75$. In addition, we observe that $\langle S_z^{\rm AFM}\rangle$ still vanishes, which actually is the result of the incommensurate spin pattern.

 \begin{figure}[tbp]
 \centering
 \includegraphics[width=3.4in]{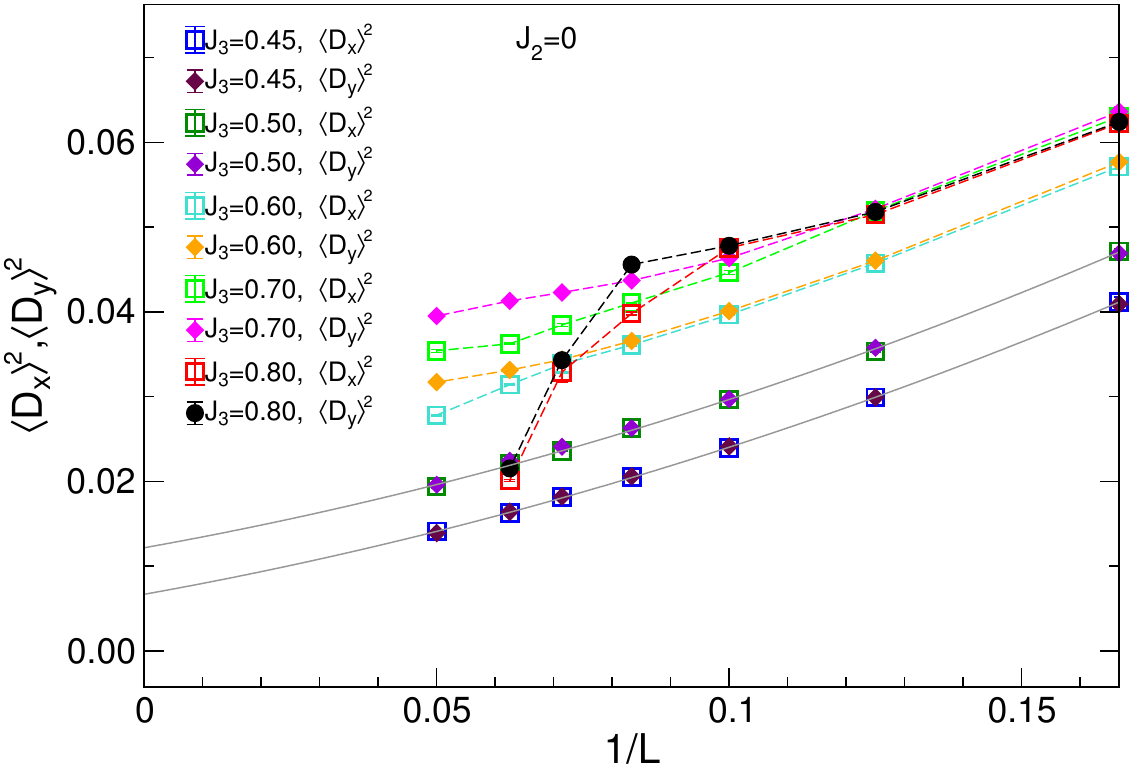}
 \caption{Dimer orders along $x-$ and $y-$ directions on $L\times L$ systems with $L=6-20$ at different $J_3$ using a fixed $J_2=0$. Solid gray lines denote second order polynomial fits for $\langle D_x\rangle^2$ at $J_3=0.45$, 0.50. See Appendix~\ref{app:fss} for detailed scaling analyses. }
 \label{fig:DimerOrder}
 \end{figure}

\subsubsection{Nature of the VBS phase}

 We also check the dimer order parameters, defined as~\cite{liuQSL}
\begin{equation}
\langle D_{\alpha}\rangle=\frac{1}{N_b}\sum_{{\bf i}}(-1)^{i_{\alpha}}\langle B^{\alpha}_{\bf i} \rangle,
\end{equation} 
where $B^{\alpha}_{\bf i}={\bf S}_{{\bf i}} \cdot {\bf S}_{{\bf i}+{\rm e_\alpha}}$ is the bond operator between site ${\bf i}$ and site ${\bf i}+{\rm e_\alpha}$ along $\alpha$ direction ($\alpha=x$ or $y$).  $N_b=L(L-1)$ is the total number of counted bonds.  As seen in Fig.~\ref{fig:16x16J1J3localOrder} of the results on a finite $16\times 16$ cluster, the VBS order parameters $\langle D_x\rangle^2$ and $\langle D_y\rangle^2$ also a show sudden drop from the VBS phase (III) to the spiral phase (IV), consistent with a first-order phase transition scenario. 

Furthermore, we observe that $\langle D_x\rangle^2$ and $\langle D_y\rangle^2$ are equal for $J_3 < 0.55$ and start to deviate for $J_3 > 0.55$, suggesting a transition from the plaquette VBS to the mixed-VBS phase.  
We believe that the difference between $\langle D_x\rangle^2$ and $\langle D_y\rangle^2$ might be induced by short-range spiral correlations (SRSC) (shown in Fig.~\ref{fig:12x28corrVBS_Spiral} and discussed later).  
By analyzing spin correlations, we find that the SRSC appears already at $J_3 \simeq 0.45$. This observation suggests that the stronger SRSC at $J_3\simeq 0.55$ may induce different VBS order parameters $\langle D_x\rangle^2$ and $\langle D_y\rangle^2$, which is a characteristic feature of the mixed-VBS state.

To examine the size dependence of the dimer order parameters, we present $\langle D_x\rangle^2$ and $\langle D_y\rangle^2$ with respect to $1/L$ in Fig.~\ref{fig:DimerOrder}. One can find clearly that at $J_3 = 0.8$ the dimer orders tend to vanish in the thermodynamic limit, in contrast to other $J_3$ cases inside the estimated VBS phase region $0.37 < J_3 \lesssim 0.75$. 
For $J_3=0.45$ and $0.5$, through finite-size scaling up to $20 \times 20$ sites, the extrapolated $\langle D_x\rangle^2$ and $\langle D_y\rangle^2$ values in the thermodynamic limit are identical, indicating a plaquette VBS order.
For $J_3=0.6$ and $0.7$, the extrapolated values of $\langle D_x\rangle^2$ and $\langle D_y\rangle^2$ are nonzero but different, suggesting a mixed columnar-plaquette VBS order.  
More details are provided in Appendix~\ref{app:fss}. 
The transition point between the plaquette and mixed VBS phase is estimated to be $J_3 \simeq 0.55$ according to the results of the $16\times 16$ size, which is shown as the violet dotted line in Fig.~\ref{fig:16x16J1J3localOrder}.

We note that the exact diagonalization of the system with twisted boundary conditions on the size up to $32$ sites, marks the region $0.5 \lesssim J_3 \lesssim 0.57$ as a mixed columnar-plaquette VBS phase, which is followed by a SRSC state before entering the long-range spiral ordered phase~\cite{ed2010}. 
This picture is roughly consistent with our results, but we believe that the SRSC state found by exact diagonalization is actually part of the VBS phase. 
In addition, a short-range valence bond study suggests a plaquette VBS state along the line $J_2+J_3=J_1/2$~\cite{PVB4}. 
In combination with our results, the VBS phase is most likely comprised of a plaquette and a mixed columnar-plaquette phase, and the transition between them is continuous. 
A similar scenario has been reported in a quantum dimer model on the square lattice, where a mixed columnar-plaquette VBS phase continuously intervenes between a columnar and a plaquette VBS phase~\cite{mixVBS}. 
Qualitatively, we know that the competition between $J_1$ and $J_3$ couplings tends to induce a spiral order. 
However, when $J_3$ is too small to lead to the SRSC, the isotropy between the $x-$ and $y-$direction survives, consistent with the existence of a plaquette VBS phase. 
When $J_3$ gets larger but still not large enough to stabilize a spiral long-range order, rotation symmetry breaks down and gives rise to a mixed columnar-plaquette VBS phase.

\subsubsection{Energy curve versus $J_3$}
\label{sec:energyJ3}
    
   \begin{figure}[htbp]
 \centering
 \includegraphics[width=3.2in]{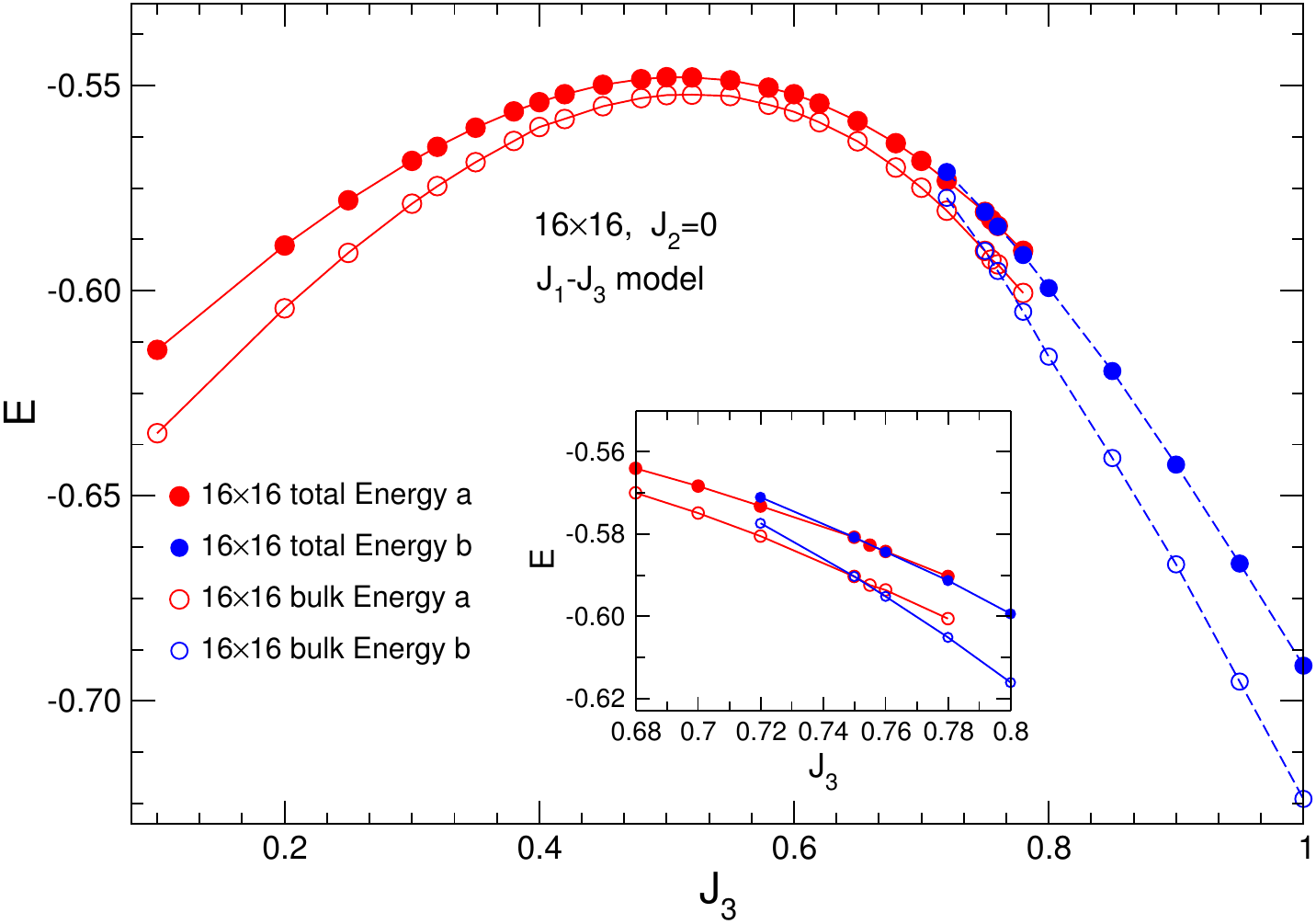}
 \caption{ Ground state energy per site at different $J_3$ ($J_1=1$ and $J_2=0$) on a $L\times L$ ($L=16$) open system using the full cluster (filled circle) or only the central $(L-8)\times (L-8)$ bulk region (unfilled circle). 
 For the cases of $J_3>0.65$ (red symbols), we use the ground state with a slightly smaller $J_3$ as an initialization for optimization. As a comparison, we also use the ground state with $J_3=0.8$ as an initialization for optimization (blue symbols). Thanks to the hysteresis, the first-order phase transition point can be located by the crossing of the two curves. The inset shows a zoom in around $J_3=0.75$ to clarify the transition point.}
\label{fig:16x16energyJ1J3model}
 \end{figure}    

  \begin{figure*}[htbp]
 \centering
 \includegraphics[width=6.2in]{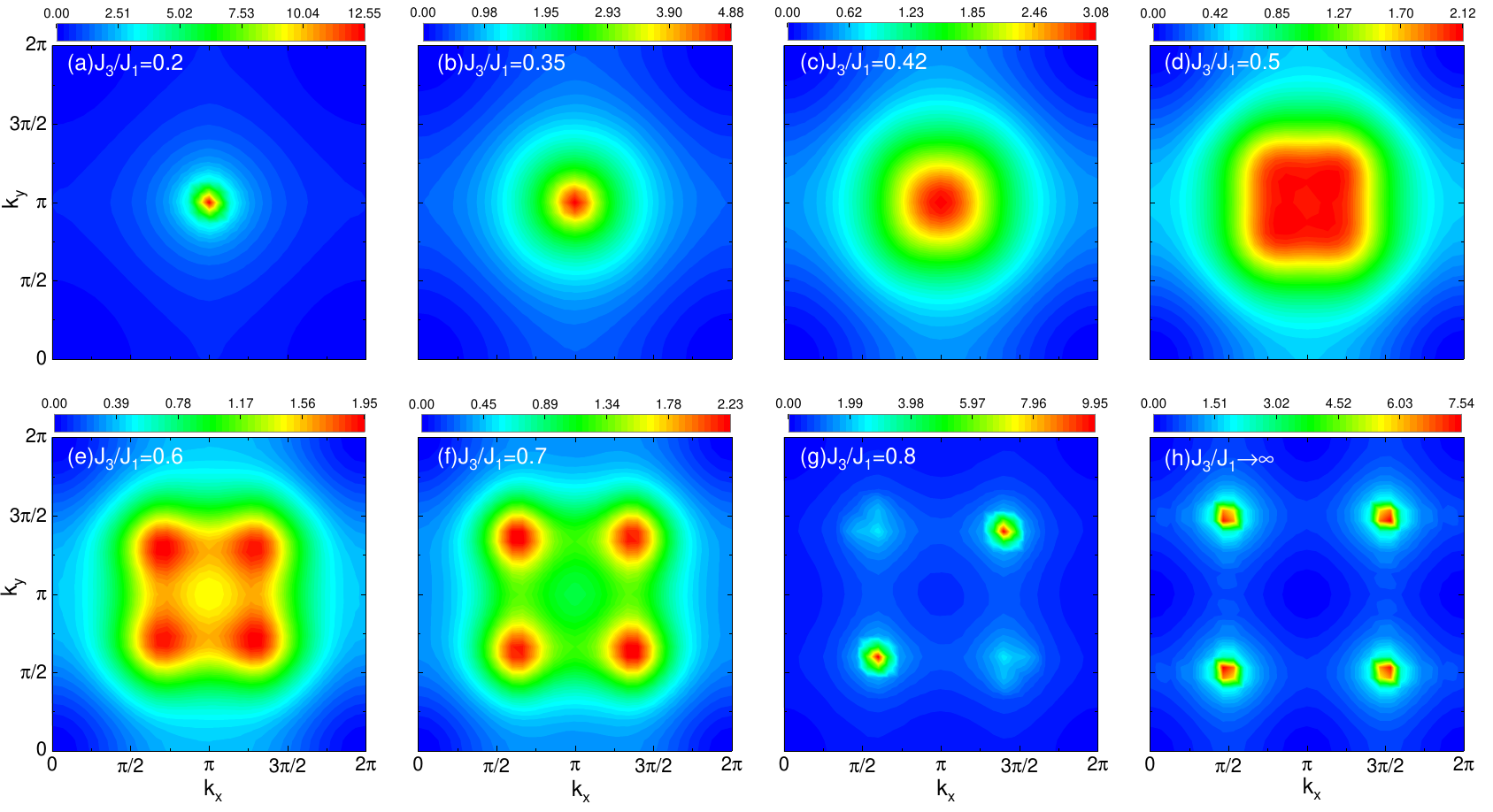}
 \caption{Given $J_2=0$,  evolution of the spin structure factor with increasing $J_3/J_1$  on $16\times 16$ sites (a)-(h).  (a) $J_3/J_1=0.2$ is in the AFM phase; (b) $J_3/J_1=0.35$ is in the gapless QSL phase; (c)-(f) $J_3/J_1=0.42,0.5,0.6,0.7$ are in the VBS phase; (g) $J_3/J_1=0.8$ is an incommensurate spiral phase with magnetic Bragg peaks at $(q,q)$ and $(-q,-q)$ where $q\simeq\frac{3}{5}\pi$; (h) $J_3/J_1\rightarrow \infty$ is a $S(\pi/2,\pi/2)$ phase. 
 }
 \label{fig:spinFactor16x16J2_0}
 \end{figure*}
      
To have a more comprehensive understanding of the various phases, we also consider the $J_3$ dependence of ground-state energy. 
In Fig.~\ref{fig:16x16energyJ1J3model}, we plot the energy per site at different $J_3$ based on the $16 \times 16$ system size. 
The filled circles (red and blue) denote the energy per site obtained from all the sites, and the empty ones (red and blue) are the energy per site obtained from the central $(L-8) \times (L-8)$ sites. 
One can see that the energy of the $16 \times 16$ system has a maximum at $J_3 \approx 0.52$, which interestingly is accompanied with the sign change of the third-nearest neighbour terms $\langle {\bf S_i}\cdot {\bf S_j}\rangle$, i.e. the third-nearest neighbour $\langle {\bf S_i}\cdot {\bf S_j}\rangle$ is positive for $J_3 \lesssim 0.52$ but negative for $J_3 \gtrsim 0.52$.
To locate the first-order transition point from the VBS to spiral phase from the ground-state energy curve, we first use the ground state at $J_3=0.65$ (with initialization by the simple update~\cite{simpleupdate}) as an initialization for further optimization to obtain the ground state  at $J_3=0.68$.
Next, we use the state at $J_3=0.68$ to get the optimized state at $J_3=0.7$, and further on to get the optimized states of $J_3=0.72$, $0.75$ and $0.78$ sequentially. 
Such a process can be viewed as an adiabatic evolution process.  
On the other hand, we can also perform the reverse process starting from the ground state at $J_3=0.8$ and move backwards sequentially to get the optimized states at $J_3=0.78$, $0.755$, $0.75$ and $0.72$.  
Except in rare special cases, first-order transitions generically show hysteresis under such an evolution, i.e. one does not adiabatically evolve into another phase.
The transition point can be obtained from the crossing of the two energy evolution curves. 
In Fig.~\ref{fig:16x16energyJ1J3model}, one can see that the energy curves of the $L \times L$ ($L=16$) system obtained from the two different paths (the red and blue solid circles) show a crossing around $J_{3}\simeq0.75$. 
Based on the energy curves of the central $(L-8) \times (L-8)$ sites which can provide a better estimation of the energy in the thermodynamic limit, the crossing is clearer. 
Both analyses indicate that the VBS-spiral transition occurs at $J_3 \simeq 0.75$, in good agreement with the behaviour of the spin order parameter $\langle S^{\rm abs}_z \rangle$ which shows an apparent discontinuity at the same  $J_3$ value, as shown in Fig.~\ref{fig:16x16J1J3localOrder}. 
We note that the VBS-spiral transition point can even be estimated via the second derivative of the energy with respect to $J_3$ on a periodic $4 \times 4$ cluster, which provides the identical result as shown in Appendix~\ref{app:ed}.

    \begin{figure}[htbp]
 \centering
 \includegraphics[width=3.4in]{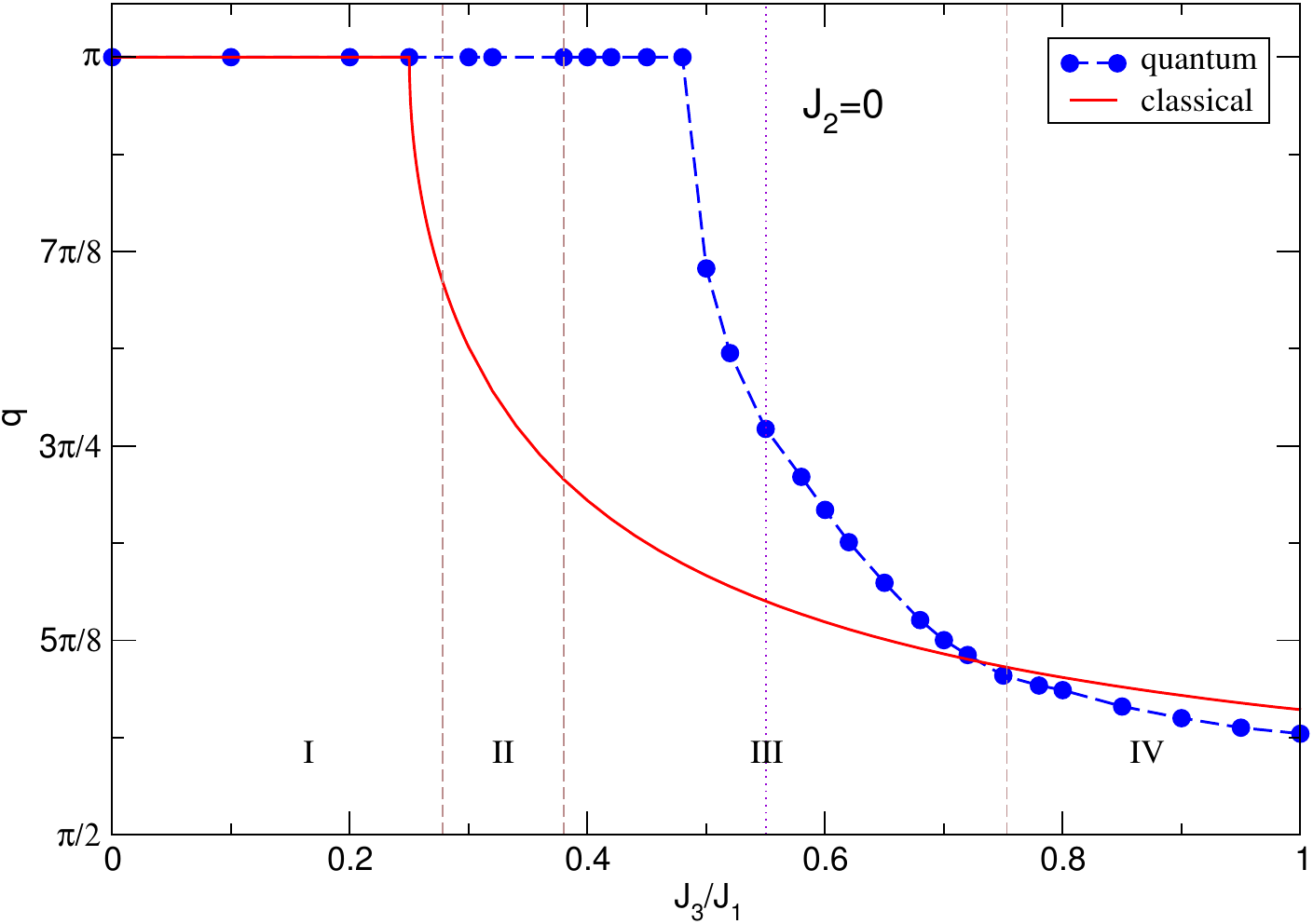}
 \caption{The variation of the peak position ${\bf Q}=(q,q)$ in the spin structure factor $S({\bf k})$. 
 Blue symbols represent the peak position obtained in the spin-1/2 $J_1$-$J_3$ model on a $16\times 16$ cluster. Red lines denote the peak position from the classical $J_1$-$J_3$ model. Vertical dashed lines separate the four phases in the quantum case: (I) AFM, (II) QSL, (III) VBS and (IV) spiral phases. The vertical dotted line in the VBS phase (III) denotes the transition between the plaquette VBS and the mixed-VBS phase.}
\label{fig:spiralpeak}
 \end{figure}

 \subsubsection{Incommensurate spiral order }   
     
  \begin{figure}[htbp]
 \centering
 \includegraphics[width=3.4in]{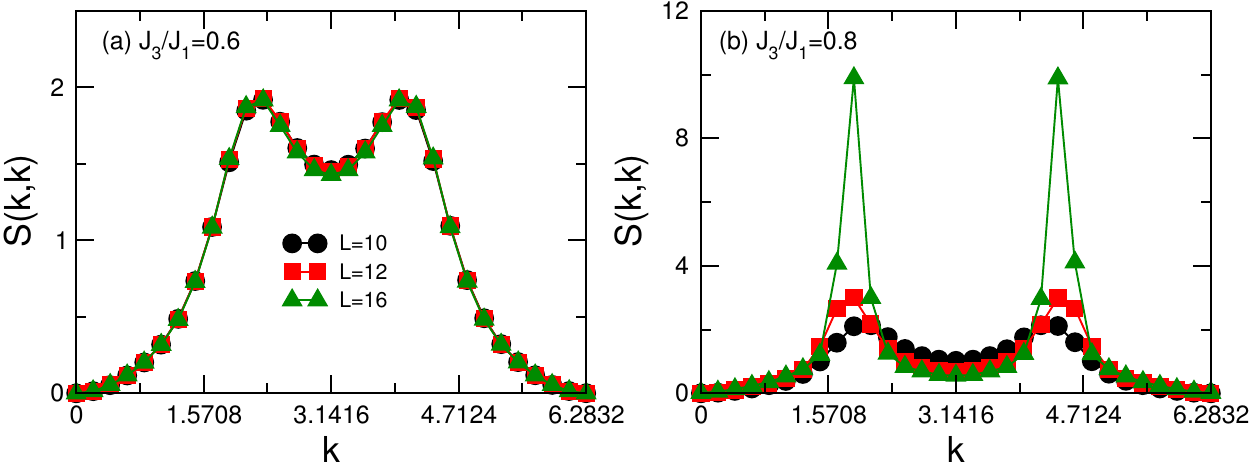}
 \caption{For a given $J_2=0$, the spin structure factor variation with system size $L$ in the $J_1-J_3$ model at $J_3=0.6$ and 0.8. $S({\bf k})=S(k_x,k_y)$ is shown along the $k_x=k_y$ line for simplicity. Clear magnetic Bragg peaks are developing at $J_3=0.8$, in contrast to $J_3=0.6$.}
 \label{fig:structureFactor}
 \end{figure}

Now we examine how the spiral order evolves with increasing $J_3$, focusing on the $J_1$-$J_3$ model, by computing the spin structure factor defined as
\begin{equation}
    S({\bf k})=\frac{1}{L^2}\sum_{\bf{ij}}\langle{\bf S}_{{\bf i}}\cdot {\bf S}_{{\bf j}}\rangle {e}^{i {\bf k}\cdot({\bf i}-{\bf j})}. 
\end{equation}
We compute all pairs of spin correlators $\langle{\bf S}_{{\bf i}}\cdot {\bf S}_{{\bf j}}\rangle$ to get the structure factor.
In the classical $J_1$-$J_3$ model, for $J_3\leq 0.25$, the ground state is a conventional N\'eel AFM state with magnetic wave vector ${\bf k_0}=( \pi, \pi)$. For  $J_3/J_1> 0.25$, the ground state has a planar incommensurate magnetic order at a wave vector ${\bf Q}=(\pm q, \pm q)$. With increasing $J_3/J_1$, the spiral order is incommensurate except at $J_3/J_1=0.5$ and $J_3/J_1 \rightarrow \infty$,  and the wave vector ${\bf Q}=(\pm q, \pm q)$ will gradually move from $q=\pi$ to $q=\pi/2$ according to the formula $q={\rm cos}^{-1}(-0.25J_1/J_3)$ for  $J_3/J_1> 0.25$~\cite{locher1990linear,classicalJ1J3}. 

  \begin{figure}[htbp]
 \centering
 \includegraphics[width=3.4in]{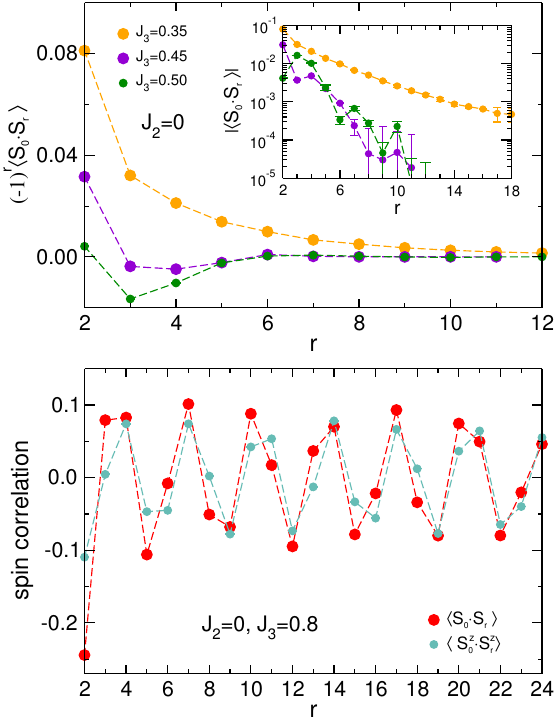}
 \caption{Spin-spin correlations along the central line $y=L_y/2$ on long strips $L_y\times L_x$ with $L_x=28$ for the $J_1$-$J_3$ model.  Upper panel: correlations at $J_3=0.45$, 0.50 (exponential decay) on $L_y=12$, compared to $J_3=0.35$ on $L_y=16$ (power-law decay discussed in \cite{liuQSL}).
 Lower panel: correlations at $J_3=0.8$ (long-range order). One sees an approximate period-10 modulation, corresponding to a wave vector $2\pi\times\frac{3}{10}=3\pi/5$. The reference site for measuring the correlations is the fourth site $(x=4)$ from the left boundary.
 }
 \label{fig:12x28corrVBS_Spiral}
 \end{figure}

\begin{figure}[htbp]
 \centering
 \includegraphics[width=3.4in]{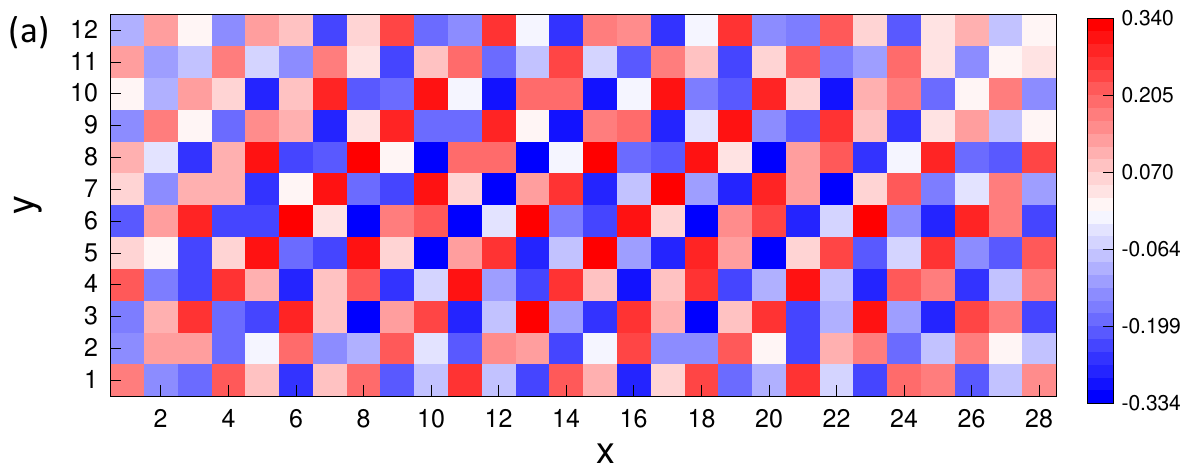}
  \includegraphics[width=3.4in]{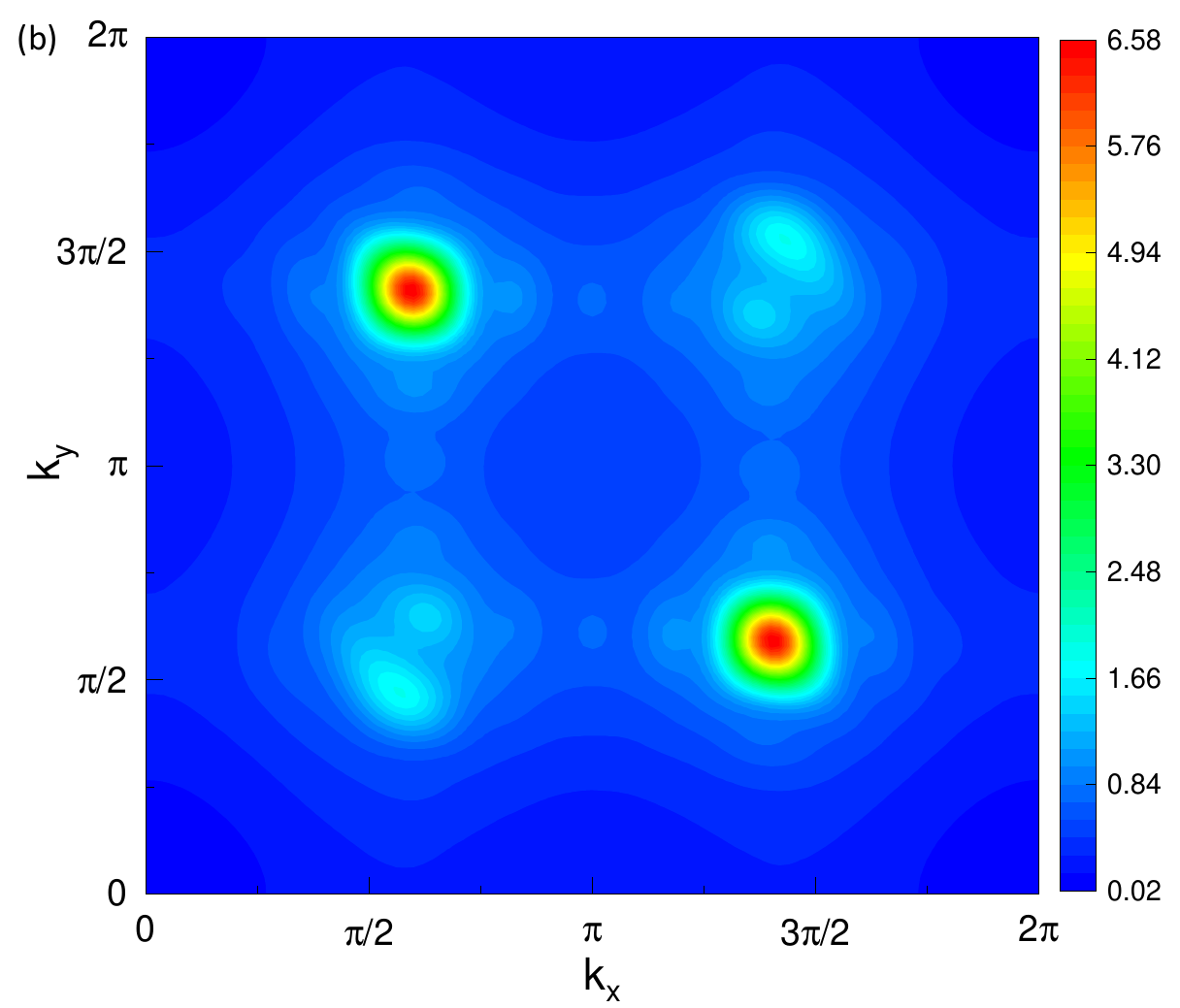}
 \caption{Spin pattern on a $12\times 28$ strip at $J_3=0.8$ $(J_2=0)$. (a) $\langle S^z_{i_x,i_y}\rangle$ distribution. (b) Spin structure factor based on $z-$component spin correlations, i.e., $S^{zz}({\bf k})$. The peak positions are located at $(q,-q)$ and $(-q,q)$ where $q\simeq\frac{3}{5}\pi$.}
 \label{fig:12x28sf}
 \end{figure}

In Fig.~\ref{fig:spinFactor16x16J2_0}, we show the spin structure factor of the $16 \times 16$ system size for different $J_3$ at $J_2=0$. At $J_3=0.2$ (in the N\'eel AFM phase), $0.35$ (in the QSL phase), and $0.42$ (in the VBS phase), the peaks of spin structure factor are located at ${\bf k_0}=(\pi,\pi)$. Further increasing $J_3$, the peak will gradually move from ${\bf k_0}=(\pi,\pi)$ to ${\bf Q}=(\pm q,\pm q)$. If $J_3/J_1\rightarrow \infty$ (set $J_1=0$), the wave vector will approach to ${\bf Q}=(\pm\pi/2,\pm\pi/2)$, as shown in Fig.~\ref{fig:spinFactor16x16J2_0}(h). In Fig.~\ref{fig:spiralpeak}, we present the classical and quantum results of the change of wave vector for a clear comparison. We would like to mention that a careful inspection of the data at $J_3 = 0.7$, as shown in Fig.~\ref{fig:spinFactor16x16J2_0}(f), reveals a detectable difference in the magnitudes of the maxima (see Appendix~\ref{app:sk}), which is consistent with the mixed VBS state breaking $\pi/2$-rotation. 

 \begin{figure}[htbp]
 \centering
 \includegraphics[width=3.4in]{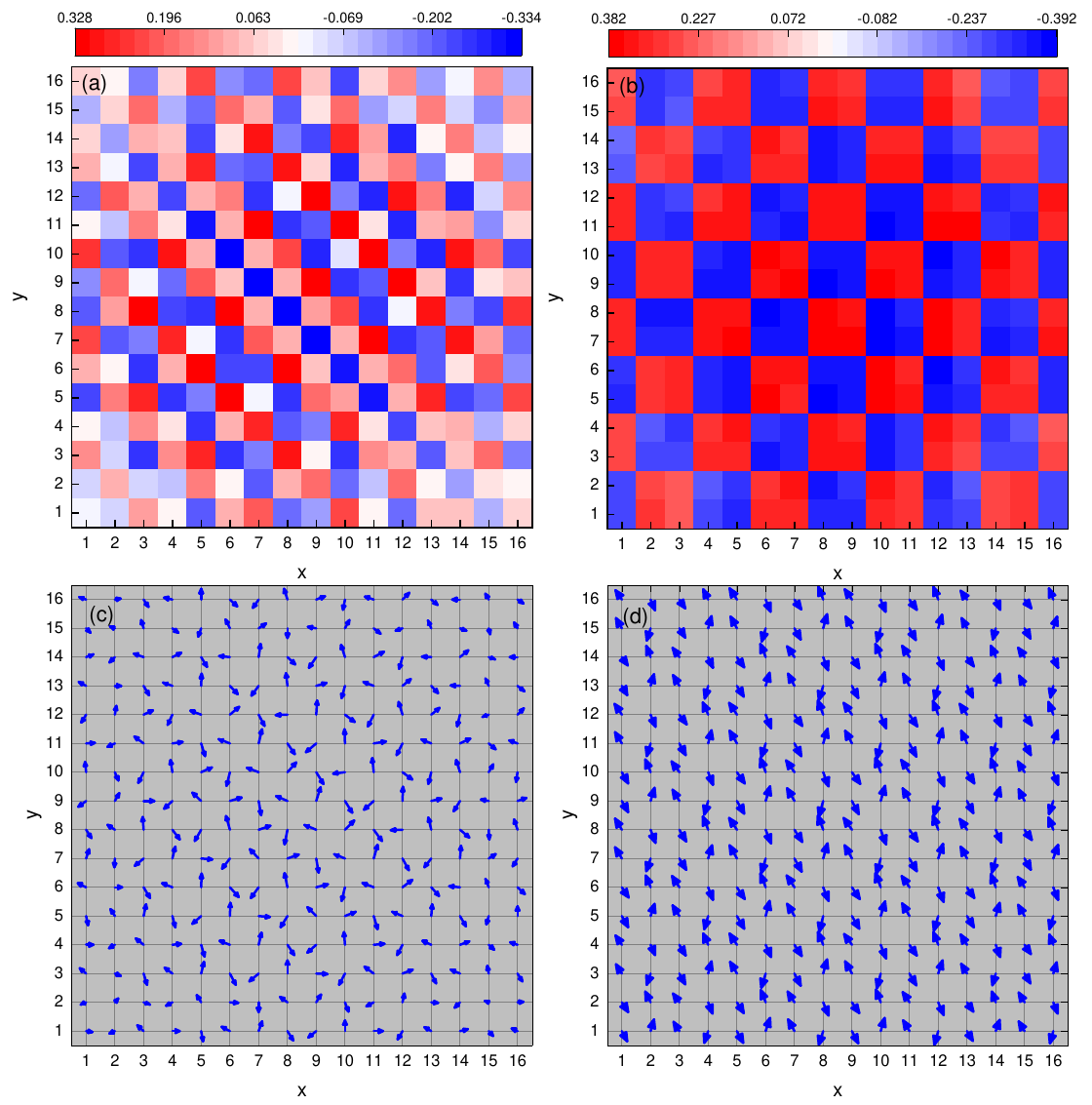}
 \caption{ Given $J_2=0$, spin patterns in the $J_1$-$J_3$ model on a $16\times 16$ open boundary system at $J_3/J_1=0.8$ (a,c), and at $J_3/J_1\rightarrow \infty$ (b,d).  (a,b) show the values of $\langle S^z_{i_x,i_y}\rangle$. (c,d) show the $\langle \Vec{S}_{i_x,i_y}\rangle$ vectors on each site in the $(x,z)$ plane, drawn as $(1.5*\langle S^x_{i_x,i_y}\rangle, 1.5*\langle S^z_{i_x,i_y}\rangle)$ in the $(x,y)$ coordinate space.}
\label{fig:spiralposition}
 \end{figure}

We point out that we find short-range incommensurate spin correlations at $J_3=0.45$ in the VBS phase, and the long-range spiral order is established for $J_3 > 0.75$, as we have discussed in the previous part. The spiral order in the VBS phase is short-ranged as the order parameter is scaled to zero in the thermodynamic limit.
As shown in Fig.~\ref{fig:structureFactor}, with increasing system size $L$, the peak value of spin structure factor at $J_3=0.6$ almost stays a constant, indicating the vanishing spin order.  
In contrast, at $J_3 = 0.8$ the peak value diverges quickly, showing the development of a spin order. 
In addition, we have also computed the spin correlation function on a $12 \times 28$ long strip for further check. In Fig.~\ref{fig:12x28corrVBS_Spiral}, the spin-spin correlations at $J_3=0.35$ (in the gapless QSL phase), $J_3=0.45, 0.5$ (in the VBS phase), and $J_3=0.8$ (in the spiral phase) are shown. In contrast to the power-law decay at $J_3=0.35$ (see more results in Ref.~\cite{liu2022emergence}), the spin correlations at $J_3=0.5$ show a clear exponential decay with an oscillation. Note that at $J_3=0.45$ and $0.5$, correlators $(-1)^r\langle S_0\cdot S_r\rangle$ have positive and negative values \footnote{There is a typo in Fig. 4(a) in  Ref.~\cite{liu2022emergence}; the y-axis label should read $|{\langle} {S_0}{\cdot} {S_r}{\rangle}|$, rather than ${(-1)^r}{\langle} {S_0}{\cdot} {S_r}{\rangle}$}. 
This is a signature of the short-range spiral order (although with a very short correlation length) while, in the long-range spiral ordered phase, the spin correlations oscillate but remain finite at long distance.

To visualize the spin pattern, we compute the values of $\langle S_{i_x,i_y}^x\rangle$, $\langle S_{i_x,i_y}^y\rangle$ and $\langle S_{i_x,i_y}^z\rangle$ on each site at $J_3=0.8$ (set $J_1=1$) and  $J_3/J_1\rightarrow \infty$ (set $J_1=0$). In Fig.~\ref{fig:12x28sf}(a) and Fig.~\ref{fig:spiralposition}, $\langle S_{i_x,i_y}^z\rangle$ on each site are presented. At $J_3=0.8$ we see a clear indication of incommensurate long-range spiral order, while for $J_3/J_1 \rightarrow \infty$ it is commensurate with a period of $4$ lattice spacings along both $x$ and $y$ directions and the 3rd-neighbour spin pairs being antiparallel. The spiral orders can be more explicitly visualized by also considering the $x$ and $y$ spin components. We find that the magnitude of the $y$-components $\langle S^y_{i_x,i_y}\rangle$ is extremely small, which is at least three orders weaker than $\langle S^z_{i_x,i_y}\rangle$ or $\langle S^x_{i_x,i_y}\rangle$ according to our resolution, indicating that the spiral order is nearly coplanar (see Figs.~\ref{fig:spiralposition}(c) and \ref{fig:spiralposition}(d)).   
The incommensurate (commensurate) long-range spiral order is also revealed by the magnetic Bragg peaks at two (four) incommensurate (commensurate) wave vectors, as shown in Fig.~\ref{fig:spinFactor16x16J2_0}(g) and Fig.~\ref{fig:12x28sf}(b)  (Fig.~\ref{fig:spinFactor16x16J2_0}(h)).

\subsection{$J_2\neq0$}

   \begin{figure}[htbp]
 \centering
 \includegraphics[width=3.4in]{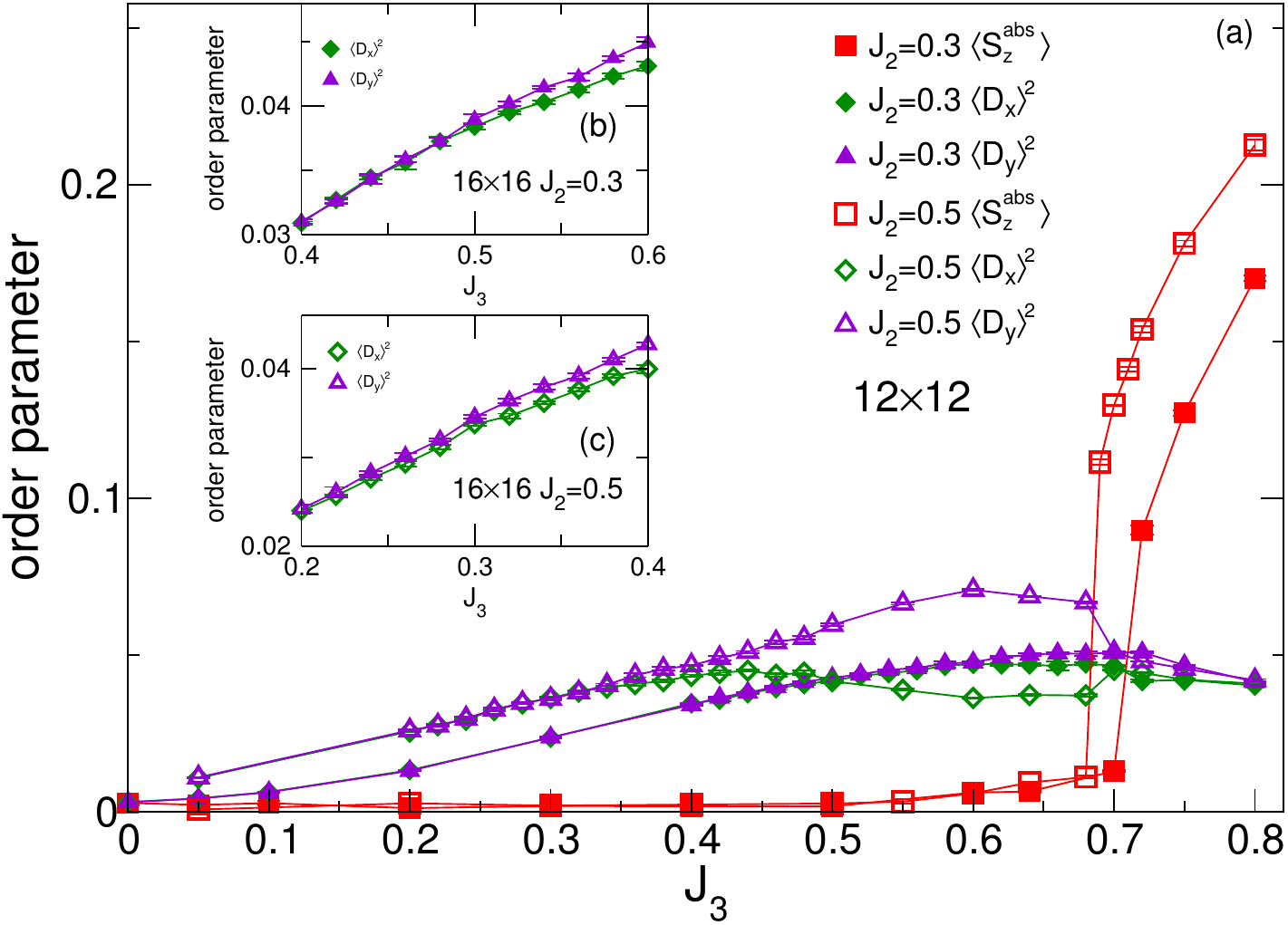}
 \caption{The $J_3 $ dependence of the spin and dimer order parameters at fixed $J_2=0.3$ and $J_2=0.5$ values on $12\times 12$ sites. For both cases $\big<S_z^{\rm AFM}\big>\simeq 0$ (not shown). The insets (b) and (c) show the dimer order parameters on $16\times 16$ sites with different $J_3$ at fixed $J_2=0.3$ and $J_2=0.5$, respectively. 
 }
 \label{fig:12x12VBS_Spiral}
 \end{figure}

    \begin{figure}[htbp]
 \centering
 \includegraphics[width=3.4in]{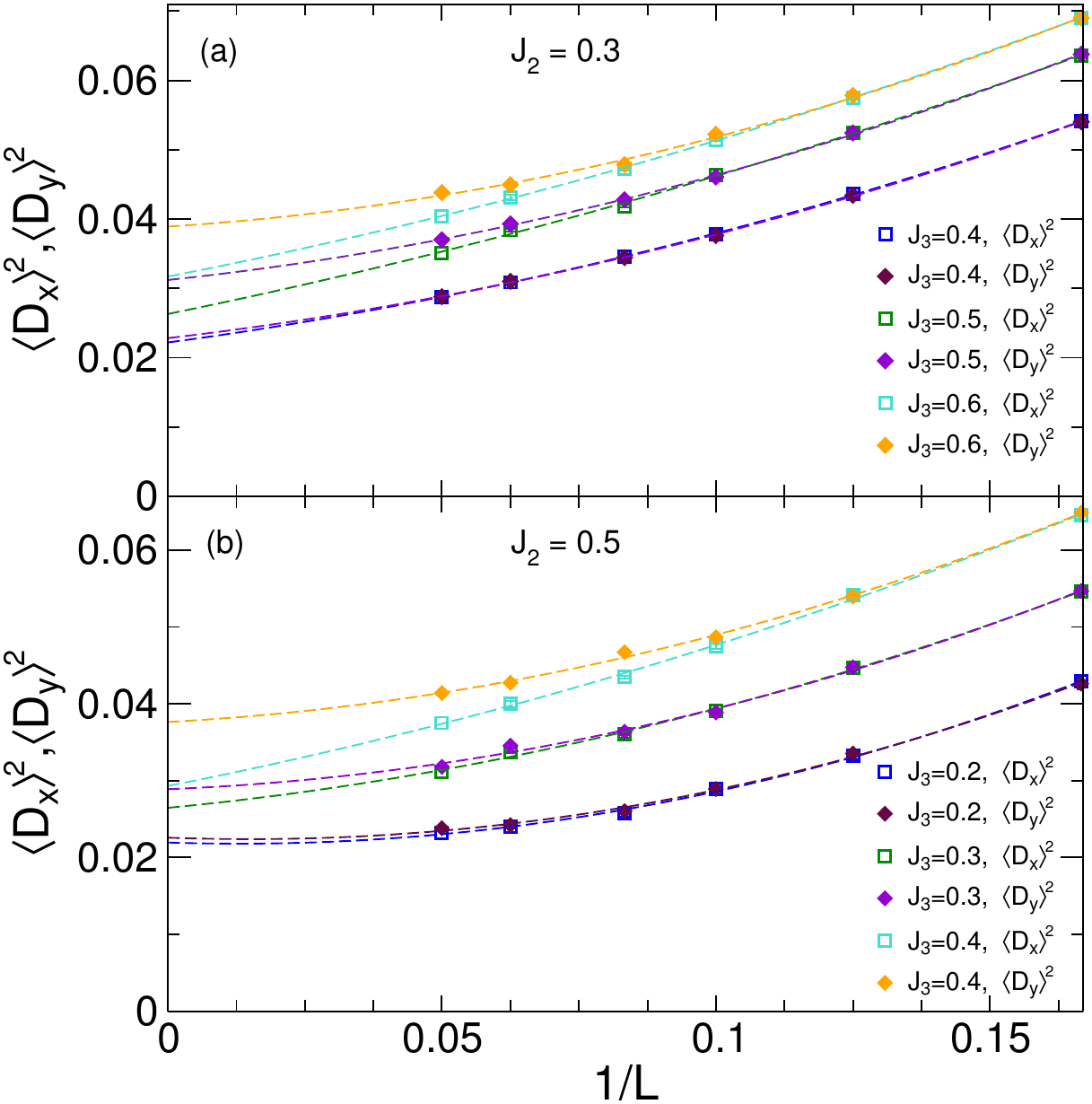}
 \caption{The $1/L$ dependence of dimer order parameters at fixed $J_2=0.3$ (a) and $J_2=0.5$ (b). Dashed lines shows second order polynomial fits with $L=6-20$. }
 \label{fig:DimerOrder_J2_03_05}
 \end{figure}
 
Based on the above results at $J_2=0$, we establish a good understanding for the properties of the VBS phase and spiral order. 
Now we turn to the $J_2\neq0$ case to explore the whole phase diagram of the spin-$1/2$ $J_1$-$J_2$-$J_3$ model.  
First of all, we focus on two typical cases at $J_2=0.3$ and $J_2=0.5$.   
At $J_2=0.3$, increasing $J_3$ from zero, the system sequentially experiences a N\'eel AFM phase, a gapless QSL phase, and a VBS phase. 
Note that for $J_2=0.5$ there is no N\'eel AFM phase for $J_3 \ge 0$.  
In both cases, when $J_3$ is large, the system always lies in the long-range spiral ordered phase.  
Here we compute the physical quantities on the $12\times 12$ cluster to check whether there exists other possible phases between the VBS and spiral order phase. 
 
In Fig.~\ref{fig:12x12VBS_Spiral}, we present the variation of the spin and dimer order parameters with increasing $J_3$. We note that at $J_2=0.3$ and $J_3=0$ (in the N\'eel AFM phase) the AFM order parameter $\langle S_z^{\rm AFM}\rangle$ on the $12\times 12$ system size is still zero, in contrast to the results on the $16 \times 16$ system at $J_2=0$. 
This is because the size $L=12$ is still a bit too small to observe the spontaneous breaking of SU(2) symmetry occurring in the N\'eel phase.  
The other spin order parameter $\langle S_z^{\rm abs}\rangle$ shows a sharp change around $J_3=0.7$ for both $J_2=0.3$ and $J_2=0.5$, indicating a first-order transition to the long-range spiral ordered phase. 
Additionally, the dimer order parameters $\langle D_x\rangle ^2$ and $\langle D_y\rangle ^2$ tend to deviate from each other when $J_3$ is large enough, and then have sudden changes at the first-order transitions.

To further investigate the properties of the VBS region, we extend our analysis to larger system sizes, including $L=16$ and $L=20$. Taking $J_2=0.3$ as an example, as shown in Fig.\ref{fig:DimerOrder_J2_03_05}(a), we observe distinct behaviors at different $J_3$: nonzero $\langle D_x\rangle ^2=\langle D_x\rangle ^2$ in the 2D limit for smaller $J_3$ values ($J_3=0.4$), and $\langle D_x\rangle ^2\neq\langle D_y\rangle ^2$ for larger $J_3$ values ($J_3=0.5$), which are consistent with the characteristics of a plaquette VBS and a mixed VBS, respectively. 
By comparing the two cases $(J_2,J_3)=(0.3,0.4)$ and $(J_2,J_3)=(0.3,0.5)$, the transition point between the two VBS states is roughly located at $J_3=0.45(5)$. Similar features in the VBS region are also found for $J_2=0.5$, with a transition point estimated at $J_3=0.25(5)$ by comparing the extrapolated thermodynamic limit results of  $(J_2,J_3)=(0.5,0.2)$ and $(J_2,J_3)=(0.5,0.3)$, shown in Fig.\ref{fig:DimerOrder_J2_03_05}(b).
In fact, the plaquette and mixed VBS state persist across most of the VBS region, and their transition line with respect to $(J_2,J_3)$ is represented as the violet line in Fig.~\ref{fig:J1J2J3phaseDiagram}. 
We remark that except aforementioned cases, for other values of $(J_2,J_3)$, the transition points (the violet points in Fig.~\ref{fig:J1J2J3phaseDiagram}) are estimated based on the results of the $16\times 16$ system.

\begin{figure*}[htbp]
 \centering
 \includegraphics[width=6.2in]{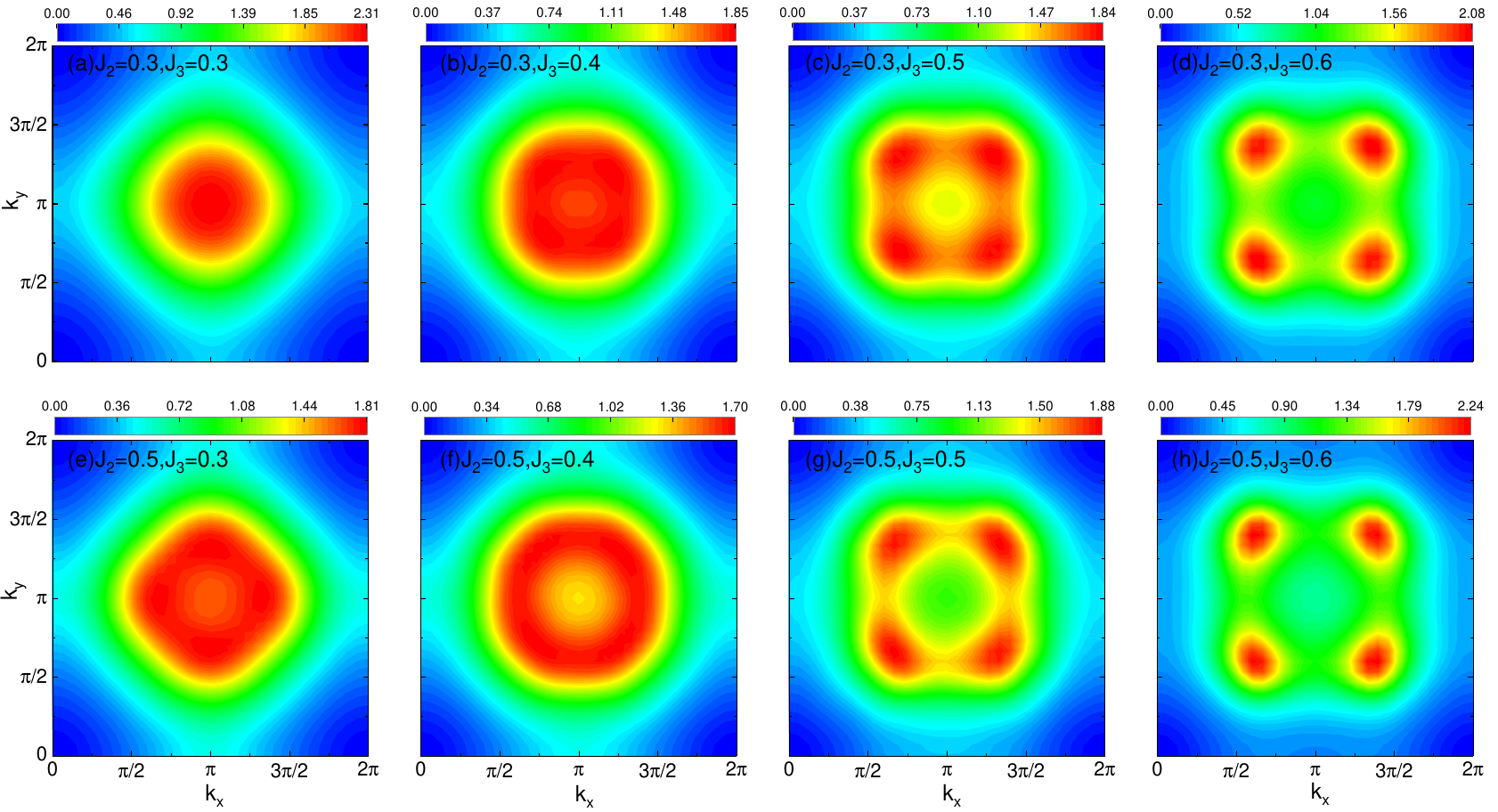}
 \caption{Spin structure factor on $12\times 12$ sites for increasing $J_3$ using fixed $J_2=0.3$ (a-d), and $J_2=0.5$ (e-h) parameters. All these points are in the VBS phase. (c,d) and (f-h) correspond to the mixed columnar-plaquette VBS phase. 
 }
 \label{fig:spinFactor12x12}
 \end{figure*}

We have also checked the evolution of spin structure factor with respect to $J_3$, as shown in Fig.~\ref{fig:spinFactor12x12}.
We find that the maximum of spin structure factor gradually moves away from $(\pi, \pi)$.  
A comparison to the results obtained at $J_2=0$ reveals the similar features for spin and dimer order parameters as well as for spin structure factor, indicating no other phases between the VBS and spiral order phase.
Interestingly, as discussed above, the emergence of short-range spiral spin correlation in some region of the plaquette VBS phase may be considered as a precursor of the transitions to the mixed VBS phase and to the spiral phase (see the transition lines reported in Fig.~\ref{fig:J1J2J3phaseDiagram}). 
 We point out that, in contrast to the classical case showing the existence of a spiral phase ordered at the wave vector $(\pi, \pm q)$ or $(\pm q, \pi)$, either our PEPS or previous exact diagonalization results do not support such a long-range ordered spiral phase in the quantum spin-$1/2$ model.

The phase diagram of the $J_1$-$J_2$-$J_3$ model is becoming more precise now: we know it must contain the N\'eel AFM, gapless QSL, VBS, stripe, and long-range spiral order phases.
Since the VBS-spiral and VBS-stripe phase transitions are both first order, we can also use the exact ground-state energies on a $4\times 4$ periodic system to evaluate the phase boundaries (see Fig.~\ref{fig:PhaseBoundaryPBC4x4} in Appendix.~\ref{app:ed}). 
We also use the PEPS results on the larger $12 \times 12$ or $16 \times 16$ systems with open boundaries to evaluate the VBS-spiral transition points for several $(J_2,J_3)$ values, showing agreement with the corresponding estimations from the periodic $4 \times 4$ cluster. The estimated phase boundaries can be seen in Fig.~\ref{fig:J1J2J3phaseDiagram}.

\subsection{AFM-spiral transition}   

It has been already shown in our previous work that for a negative $J_2\lesssim -0.25$, there exists an AFM-VBS transition line~\cite{liu2022emergence}. 
Here, we find that for a larger negative $J_2$ such as $J_2=-1.2$, the VBS phase will disappear and a direct first-order AFM-spiral phase transition occurs.

\begin{figure}[htbp]
 \centering
 \includegraphics[width=\columnwidth]{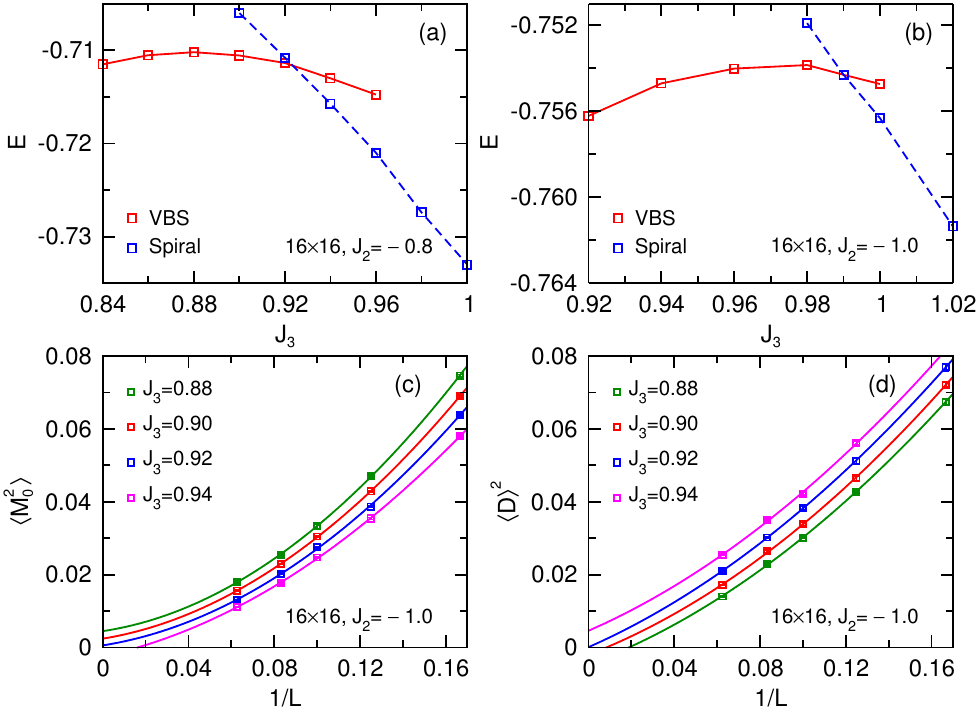}
 \caption{ Energy as a function of $J_3$ on $16\times 16$ at fixed $J_2=-0.8$ (a) and $J_2=-1.0$ (b). Red and blue symbols denote energies initialized by VBS and spiral states. (c-d) Finite size scaling of the AFM order parameter $\langle M^2_0 \rangle$ and the VBS order parameter $\langle D \rangle^2$ at different $J_3$ for a fixed $J_2=-1.0$.}
\label{fig:16x16J2_m0.8_m1.0}
 \end{figure}

We first consider the intermediate negative $J_2$ values, say $J_2=-0.8$ or $J_2=-1.0$. 
In such cases, increasing $J_3$ will lead to a continuous AFM-VBS transition followed by a first-order VBS-spiral transition. 
As discussed in Sec.~\ref{sec:energyJ3}, the VBS-spiral transition point can be generically determined by comparing the energies of the simulations with different initial states. 
In Figs.~\ref{fig:16x16J2_m0.8_m1.0}(a) and \ref{fig:16x16J2_m0.8_m1.0}(b), we show the energy variation with respect to $J_3$ on the $16\times 16$ size when a VBS-spiral transition occurs, at fixed $J_2=-0.8$ and $J_2=-1.0$ respectively. 
The AFM-VBS transition point can be evaluated by the finite-size scaling of the corresponding order parameters. 
In Figs.~\ref{fig:16x16J2_m0.8_m1.0}(c) and \ref{fig:16x16J2_m0.8_m1.0}(d), we present the size scaling of the N\'eel AFM order parameter $\langle M^2_0 \rangle$ and VBS order parameter $\langle D \rangle^2=\langle D_x \rangle^2+\langle D_y \rangle^2$ for $J_2=-1.0$ as an example, where $\langle M^2_0 \rangle = \frac{1}{L^2}S({\bf k_0})$ with ${\bf k_0}=(\pi,\pi)$. 
We find that the VBS phase is located in the interval $0.92\lesssim J_3 \lesssim 0.99$ for $J_2=-1.0$. 
Upon further increasing the magnitude of the negative $J_2$, the extension of the VBS region with $J_3$ shrinks rapidly, leading to an end point of the continuous AFM-VBS transition.

\begin{figure}[htbp]
 \centering
 \includegraphics[width=\columnwidth]{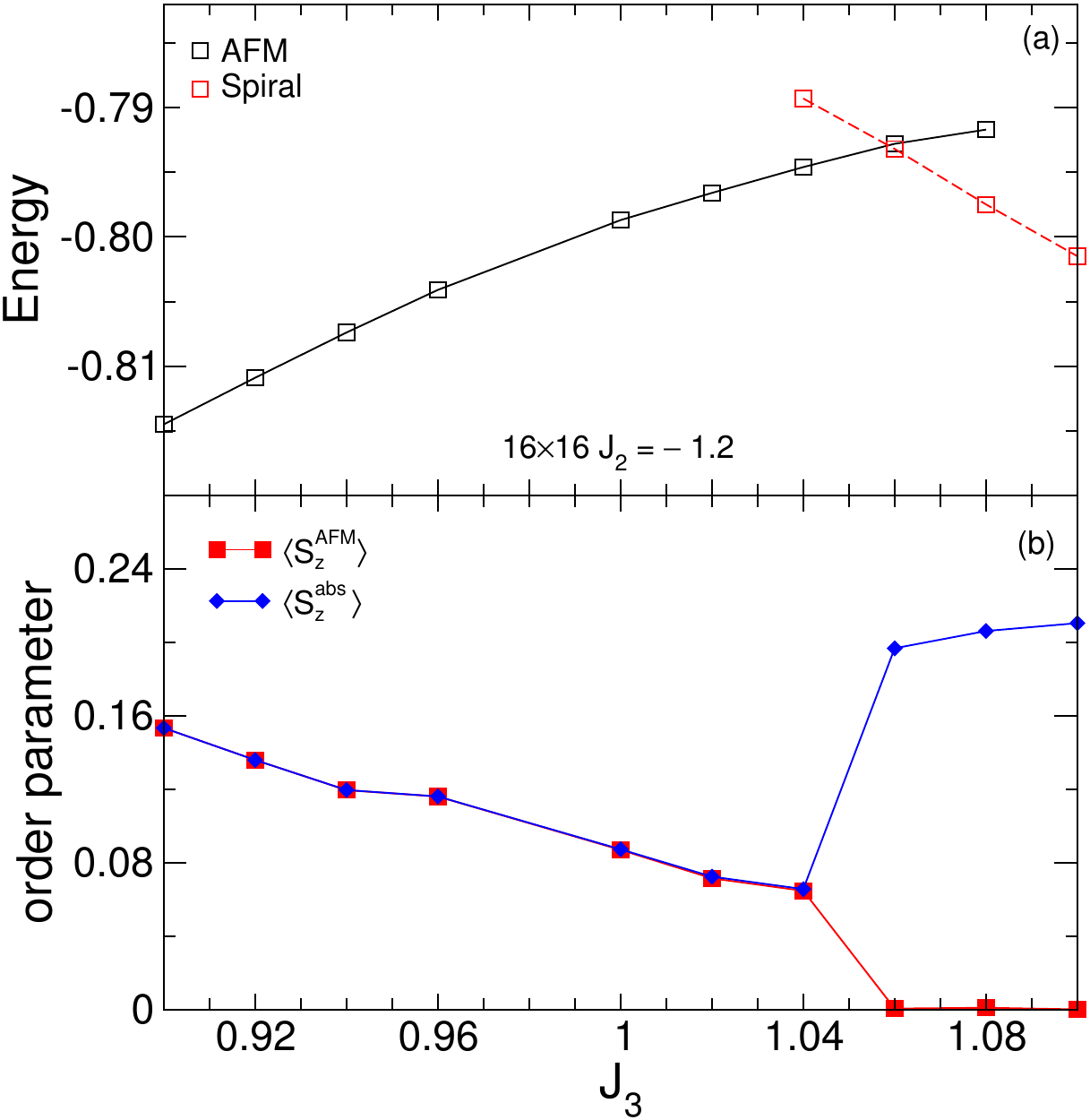}
 \caption{Energy and spin order parameters versus $J_3$ on $16\times 16$ at fixed $J_2=-1.2$.  (a) Black and red symbols denote the optimal AFM energies and optimal spiral energies, respectively. (b) The variation of the ground state  spin order parameters.  }
\label{fig:16x16J2_m1.2}
 \end{figure}

Now we turn to $J_2=-1.2$, at which we find a direct first-order AFM-spiral transition with increasing $J_3$. 
As shown in Fig.~\ref{fig:16x16J2_m1.2}, we can see the transition happening at $J_3\simeq 1.06$. 
At $J_3=1.06$, the optimal AFM state has an energy $E=-0.79285$, very close to the optimal spiral state energy $E=-0.79342$ (see Fig.~\ref{fig:16x16J2_m1.2}(a)), but they have clear AFM and spiral spin patterns correspondingly (not shown here). 
The ground-state spin order parameters also show sharp changes around $J_3=1.06$ in Fig.~\ref{fig:16x16J2_m1.2}(b), providing strong evidence for a first-order phase transition.

\subsection{$J_3<0$}

To obtain more information of the phase diagram, we finally consider the region with a negative $J_3$. 
For $J_3=0$, i.e. the $J_1$-$J_2$ model, a gapless QSL and a VBS phase emerge between the N\'eel AFM and stripe AFM phases. 
A negative $J_3$ will enhance spin orders including the N\'eel and stripe AFM orders, which therefore destabilizes the QSL and VBS phases. 
Here, we try to estimate how large a (negative) $J_3$ will be able to suppress the QSL and VBS phases.

\begin{figure}[htbp]
 \centering
 \includegraphics[width=3.4in]{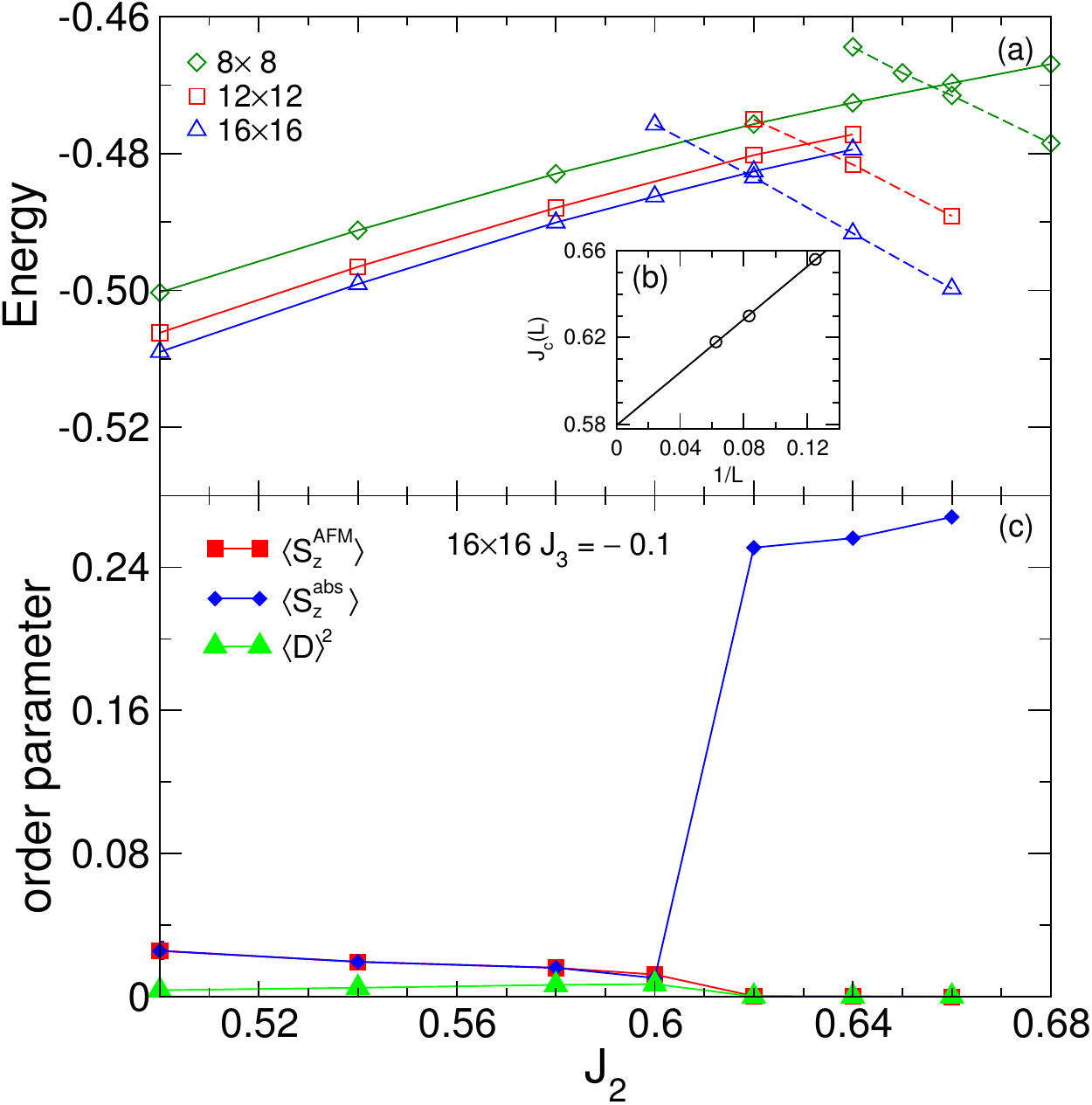}
 \caption{(a) Energy dependence versus $J_2$ on $8\times 8$, $12\times 12$ and $16\times 16$ open systems at fixed $J_3=-0.1$. The dashed lines denote the energies using the stripe state as an initialization. (b) The first-order transition point $J_2^{c}(L\rightarrow \infty)$ evaluated in the thermodynamic limit from a linear extrapolation in $1/L$ of the transition points $J_2^{c}(L)$. (c) The spin and dimer local order parameters on $16\times 16$ at different $J_2$ with fixed $J_3=-0.1$.
 }
 \label{fig:16x16J3_m0.1}
 \end{figure}

We first consider a relatively larger negative $J_3$, $J_3=-0.1$. 
In this case, we find a direct first-order transition between the N\'eel AFM and stripe phases. 
At $J_2=0.50$, we observe a clear N\'eel AFM pattern and the averaged local moment $\langle S^{z}_{i_x,i_y}\rangle$ is about $0.026$ on the $16\times 16$ cluster. 
Further increasing $J_2$ to $0.58$, this averaged moment is reduced to $0.016$, still showing a clear N\'eel AFM order.
Meanwhile, the local dimer order $\langle D \rangle^2=\langle D_x \rangle^2+\langle D_y \rangle^2$ remains much smaller, which is about $0.0067$. 
The N\'eel-stripe is a typical first-order transition, and the $J_2$ transition point is expected to shift as the system size increases, similarly to the VBS-stripe transition in the $J_1$-$J_2$ model~\cite{liuQSL}. 
As discussed in Sec.~\ref{sec:energyJ3}, we can evaluate the transition point by initializing the PEPS optimization from either the N\'eel or the stripe state.
Using the crossing of the energy curves shown in Fig.~\ref{fig:16x16J3_m0.1}(a), the transition points $J_2^c(L)$ can be obtained as $0.656$, $0.630$, and $0.618$ for $8\times8$, $12\times 12$ and $16\times 16$, respectively.
Furthermore, we take a linear extrapolation of $J_2^c(L)$ versus $1/L$, giving the first-order transition point in the thermodynamic limit $J_2^{c}(L\rightarrow\infty) \simeq 0.58$, as shown in Fig.~\ref{fig:16x16J3_m0.1}(b).
As the bulk energy on the $16\times 16$ size can better approximate the energy in the thermodynamic limit, following Ref.~\cite{liuQSL}, we also use the central $4\times 4$ bulk energy of the $16\times 16$ open system to estimate the thermodynamic limit transition point, which gives a consistent value $J_2^{c}(L\rightarrow\infty) \simeq 0.582$ (not shown here). 
We also demonstrate the spin and dimer order parameters on the $16 \times 16$ size, as shown in Fig.~\ref{fig:16x16J3_m0.1}(c), which also confirms the first-order transition nature.

We further consider a smaller $J_3$, say, $J_3=-0.05$. 
By taking $J_2=0.58$, we find that the spin order $\langle S^z_{i_x,i_y}\rangle$ at $(J_2,J_3)=(0.58, -0.05)$ is $0.008$, which is half of the value obtained at $(J_2,J_3)=(0.58, -0.1)$. 
This result indicates that $(J_2,J_3)=(0.58, -0.05)$ could still be in the N\'eel AFM phase, but rather close to the nonmagnetic regime.
These results suggest that both the gapless QSL and VBS phases may disappear at a quite small negative value of $J_3$. 
In such a narrow region, it is hard to accurately determine the transition lines of the QSL and VBS phases. 
Therefore, we schematically show in the inset of Fig.~\ref{fig:J1J2J3phaseDiagram} the phase boundaries of the QSL and VBS phases for $J_3<0$ ending in two multicritical points.  
Here we assume that the gapless QSL does not touch the stripe phase but remains separated by the VBS phase. 
We believe that a first-order transition always occurs between the stripe phase and other phases, in that case the AFM and VBS phases. As the two putative multicritical points are very close, another possible scenario is that the N\'eel AFM, QSL, VBS and stripe phases are directly connected by a quadruple point.

\section{Conclusion and discussion} 

In this work, we establish the global phase diagram of the square-lattice $J_1$-$J_2$-$J_3$ model using finite PEPS simulations with careful finite-size scaling. 
We compute spin and dimer order parameters, as well as spin-spin correlation functions. 
First, we focus on the identification of the VBS phases and find a novel transition from the 4-fold degenerate plaquette VBS phase (beyond the boundary with the previously discovered QSL phase) to a 8-fold degenerate mixed-VBS phase (e.g. in the $J_1$-$J_3$ model with increasing $J_3$). 
The mixed-VBS phase can be viewed as spontaneously breaking the point group $C_4$ symmetry of the plaquette VBS state.
In other words, while the horizontal and vertical dimers have the same magnitude in the plaquette VBS phase, they start to become different at the transition to the mixed-VBS. 
The mixed-VBS phase also shows incommensurate short-range spin correlations while approaching the magnetic spiral phase ordered at wave vectors $(\pm q, \pm q)$. 
Our results combined with previous studies suggest that the transition between the plaquette and mixed columnar-plaquette VBS phases is continuous as expected from the Ginzburg-Landau paradigm and studies of quantum dimer models~\cite{mixVBS}. 
In contrast, the transition between the mixed-VBS and spiral phase appears to be first order. 
In the spiral phase, we establish the existence of long-range spiral spin correlation and explicitly visualize the incommensurate spiral patterns in real space.

Therefore, the overall ground-state phase diagram of the $J_1$-$J_2$-$J_3$ model is elucidated in detail, which contains six phases: a N\'eel AFM phase ordered at $(\pi,\pi)$, a stripe AFM phase ordered at $(\pi,0)$ or $(0,\pi)$, two VBS phases, a gapless QSL phase,  and a spiral phase ordered at $(\pm q, \pm q)$ including the state ordered at $(\pm \pi/2, \pm \pi/2)$.
We would like to stress that our work significantly broadens the knowledge on the VBS and spiral order phase. 
It also provides a canonical example for the understanding of quantum effects and competition caused by frustration. Interestingly, we find that the QSL and VBS phases in the $J_1$-$J_2$ model, which are easily suppressed by a very small negative $J_3 = -0.02\sim -0.05$, are rather close to the multicritical points occurring at a quite small $J_3<0$, as seen in Fig.~\ref{fig:J1J2J3phaseDiagram}. 
This closeness may naturally explain the very long correlation lengths found  in the previous studies of the pure $J_1$-$J_2$ model for the nonmagnetic region~\cite{gong2014,poilblanc2017,liuQSL}.

Furthermore, we find two distinct types of multicritical points in the phase diagram. 
The first type involves the critical points at which three continuous transition lines intersect: AFM-VBS, AFM-QSL, and QSL-VBS (marked as the two blue dots in Fig.~\ref{fig:J1J2J3phaseDiagram}). 
The second type encompasses the critical points that mark the termination of a continuous line culminating in a first-order transition, which occurs with the AFM-VBS critical line reaching either the spiral phase or the stripe phase (denoted as the two red dots in Fig.~\ref{fig:J1J2J3phaseDiagram}). It is noteworthy that these two types of critical points are conceptually different. Intriguingly, the continuous AFM-VBS transition functioning as a line of deconfined quantum critical points (DQCP)~\cite{liu2022emergence}, can culminate at both types of multicritical points. 
This observation can offer valuable insights into comprehending the nature of the DQCP~\cite{DQCP1,DQCP2}.

Finally, we would like to point out that our study further demonstrates the capability of finite PEPS as a powerful numerical tool to study strongly correlated 2D quantum many-body systems.  
Since the tensor elements of finite PEPS can be independent, the finite PEPS constitutes a very general ansatz family and can naturally capture nonuniform properties.  
In particular, the power of finite PEPS has been fully explored when it is used to accurately represent incommensurate short-range or long-range orders.
We note that in fermionic correlated systems like the $t-J$ model, short-range incommensurate correlations can also exist~\cite{spinwave1990}, and hence the finite PEPS method should be able to provide accurate results for such systems.

\section{Acknowledgment}    
      
We thank Gang Chen,  Zheng-Xin Liu,  Han-Qiang Wu and Rong Yu for helpful discussions. 
This work is supported by the CRF C7012-21GF, the ANR/RGC Joint Research Scheme No. A-CUHK402/18 from the Hong Kong's Research Grants Council and the TNTOP ANR-18-CE30-0026-01 grants awarded from the French Research Council. Wei-Qiang Chen is supported by the National Key R\&D Program of China (Grants No. 2022YFA1403700),  NSFC (Grants No. 12141402), the Science, Technology and Innovation Commission of Shenzhen Municipality (No. ZDSYS20190902092905285), Guangdong Basic and Applied Basic Research Foundation under Grant No. 2020B1515120100, and Center for Computational Science and Engineering at Southern University of Science and Technology. S.S.G. was supported by the NSFC (No. 12274014), the Special Project in Key Areas for Universities in Guangdong Province (No. 2023ZDZX3054), and the Dongguan Key Laboratory of Artificial Intelligence Design for Advanced Materials (DKL-AIDAM).  W.Y.L. was  supported by the U.S. Department of Energy, Office of Science, National Quantum Information Science Research Centers, Quantum Systems Accelerator. 

  \vspace{0.4cm}    
  
\appendix
\section{convergence with $D$}
\label{app:converge}

Here we  check the $D$-convergence on a $16\times 16$ open system at $(J_2,J_3)=(-1, 0.8)$. In Table.\ref{tab:convergence}, we present the obtained $D=4-10$ results for energy, dimerization and magnetization. It clearly indicates $D=8$ is sufficient to get converged results.

\begin{table}[htb]
   \centering
 \caption { The convergence of physical quantities with respect to the PEPS bond dimension $D$ on a $16\times 16$ system at $(J_2,J_3)=(-1, 0.8)$, including  ground state energy per site, dimerization $\langle D \rangle^2=\langle D_x \rangle^2+\langle D_y \rangle^2$, and AFM magnetization $\langle M^2_z \rangle=\langle (S^{\rm AFM}_z)^2\rangle$ (see main text for detailed definitions). }
	\begin{tabular*}{\hsize}{@{}@{\extracolsep{\fill}}cccc@{}}
		\hline\hline
	   $D$  & $E$ & $\langle D \rangle^2 $    & $\langle M^2_z \rangle$    \\ \hline
 		
        4 & -0.769042(6) & 0.00656(9) & 0.0290(1)  \\
        6 & -0.772245(4) & 0.00686(8) & 0.0261(2) \\
        8 & -0.773084(3) & 0.00701(6) & 0.0217(1) \\  
        10 &-0.773092(5) & 0.00701(8) & 0.0218(2)\\  
 		\hline\hline
	\end{tabular*}
\label{tab:convergence}	
\end{table}

\section{Finite size scaling of VBS order parameters}
\label{app:fss}

\begin{figure}[tbp]
 \centering
 \includegraphics[width=3.4in]{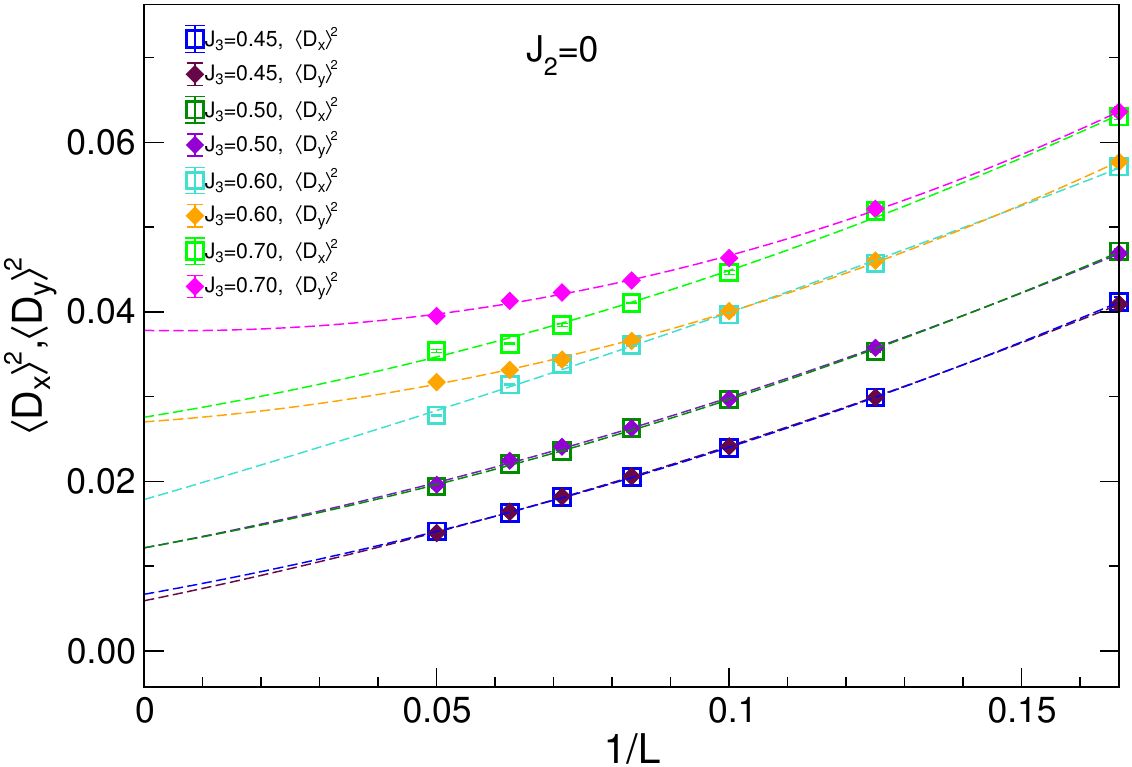}
 \caption{Dimer orders along $x-$ and $y-$ directions on $L\times L$ systems with $L=6-20$ at different $J_3$ using a fixed $J_2=0$. Second order polynomial fittings are used. }
 \label{fig:DimerOrder_detailed}
 \end{figure}

Figure ~\ref{fig:DimerOrder_detailed} presents the $1/L$ dependence of the dimer order parameters $\langle D_x \rangle^2$ and $\langle D_y \rangle^2$ for the $J_1-J_3$ model (i.e. $J_2$=0), with different $J_3$ values inside the VBS phase. We can see  $\langle D_x \rangle^2$ and $\langle D_y \rangle^2$ are almost identical on each size for smaller $J_3$, and their extrapolated values for 2D limit are 0.0066(5) and 0.0059(4) for $J_3=0.45$, 0.0122(8) and 0.0121(7) for $J_3=0.5$, correspondingly. This indicates  $J_3=0.45$ and $J_3=0.5$ have a plaquette VBS. While for $J_3=0.6$ and 0.7,  $\langle D_x \rangle^2$ and  $\langle D_y \rangle^2$ get gradually different with system size $L$ increasing. Furthermore, The extrapolated values of  $\langle D_x \rangle^2$ and $\langle D_y \rangle^2$ for 2D limit both are nonzero, 0.0178(16) and 0.0270(7) for  $J_3=0.6$, and 0.0276(21) and 0.0378(9) for  $J_3=0.7$, consistent with a mixed VBS phase.

\section{$4\times 4$ results from exact diagonalization}
\label{app:ed}
\begin{figure}[htbp]
 \centering
 \includegraphics[width=3.2in]{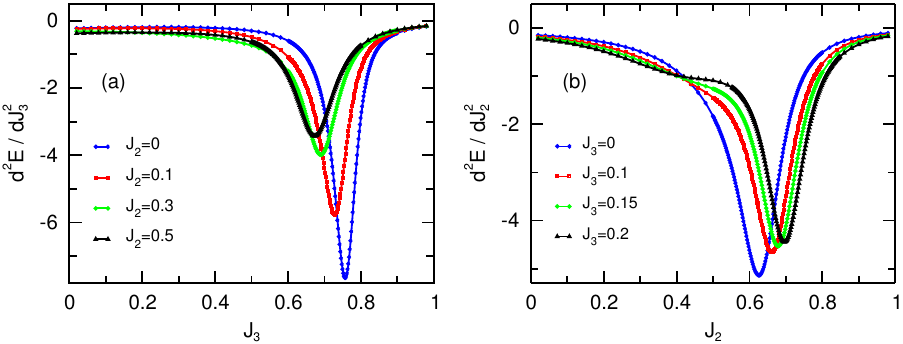}
 \caption{ (a) Second derivative of the ground state energy with respect to $J_3$ for given $J_2$ values on a periodic $4\times 4$ system  to estimate the VBS-spiral phase transition point. (b) Second derivative of the ground state energy with respect to $J_2$ for given $J_3$ values on a periodic $4\times 4$ system to estimate the VBS-stripe phase transition point.}
\label{fig:PhaseBoundaryPBC4x4}
 \end{figure}

As the VBS-spiral and VBS-stripe phase transitions are first-order, we  use the exact  ground state energies on a $4\times 4$ periodic system to estimate the phase boundaries, as shown in Fig.~\ref{fig:PhaseBoundaryPBC4x4}.  For the VBS-stripe transition, given $J_3=0$,  we have found that the  transition point is located at $J_2=0.61$~\cite{liuQSL}.  Here the second derivative of the $4\times 4$ energy with respect to $J_2$ gives an estimate of the transition point around $J_2=0.626$, in  good agreement. For the VBS-spiral transition at fixed $J_2=0$, 0.3 and 0.5,  the transition points obtained by PEPS calculations on open $16\times 16$ or $12\times 12$ systems, are $J_3\simeq 0.75$, 0.70 and 0.68, respectively, also consistent with the corresponding estimations from the periodic $4 \times 4$ cluster, namely $J_3\simeq0.757$, 0.691 and 0.674. 

\section{spin structure factor in mixed-VBS phase}

\label{app:sk}

\begin{figure}[htbp]
 \centering
 \includegraphics[width=3.4in]{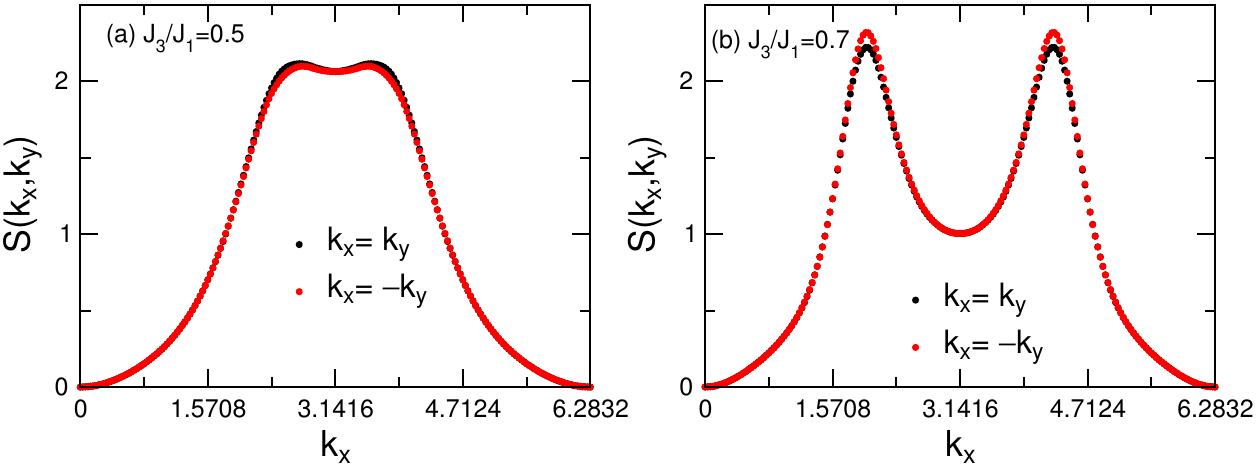}
 \caption{Spin structure factor in the $J_1$-$J_3$ model at $J_3=0.5$ and 0.7 on $16\times 16$, along $k_x=k_y$ and $k_x=-k_y$.}
 \label{fig:spinStructureFactor_kx=ky}
 \end{figure}

We have shown the contour plots of the spin structure factor $S(\bf k)$ of the $J_1$-$J_3$ model in Fig.~\ref{fig:spinFactor16x16J2_0}. Here we have a much closer look at $S(\bf k)$ in the mixed-VBS phase searching for (weak) signatures of the spontaneously breaking of the $\pi/2$ rotation symmetry. In Fig.~\ref{fig:spinStructureFactor_kx=ky}, we present $S(\bf k)$ along $k_x=k_y$  and $k_x=-k_y$ at $J_3=0.5$ (inside the plaquette VBS phase) and $J_3=0.7$ (inside the mixed columnar-plaquette VBS phase). At $J_3=0.5$, the two curves are almost identical, consistent with a $\pi/2$ rotation symmetry for the plaquette VBS. In contrast, at $J_3=0.7$, they show visible differences around the maxima, indicating the breaking of $\pi/2$ rotation symmetry in the mixed VBS phase. Note that a clear incommensurability and a larger spin correlation length (estimated from the inverse of the width of the peaks) are seen in this case.

\clearpage

\bibliography{j1j2j3longpaper}

\end{document}